\documentclass[reprint,aps,prl,superscriptaddress]{revtex4-2}
\usepackage{graphicx}
\usepackage{dcolumn}
\usepackage{bm}
\usepackage[usenames]{color}
\usepackage{xcolor}
\usepackage{tabularx}
\usepackage{booktabs}
\usepackage{hyperref}
\usepackage{siunitx}
\usepackage[utf8]{inputenc}
\usepackage[english]{babel}
\hypersetup{colorlinks,linkcolor=red,citecolor=blue,urlcolor=blue,final}
\usepackage{amsmath, amsthm, amscd, amssymb} 

\usepackage{comment}    
\usepackage[mathlines]{lineno}

\begin{document}
\title{Cooling of an optically levitated nanoparticle via measurement-free coherent feedback}
\author{Bruno Melo}
\affiliation{Nanophotonic Systems Laboratory, Department of Mechanical and Process Engineering, ETH Zurich, 8092 Zurich, Switzerland}
\affiliation{Quantum Center, ETH Zurich, 8083 Zurich, Switzerland}

\author{Dani{\"e}l Veldhuizen}
\affiliation{Nanophotonic Systems Laboratory, Department of Mechanical and Process Engineering, ETH Zurich, 8092 Zurich, Switzerland}
\affiliation{Quantum Center, ETH Zurich, 8083 Zurich, Switzerland}

\author{Grégoire F. M. Tomassi}
\affiliation{Nanophotonic Systems Laboratory, Department of Mechanical and Process Engineering, ETH Zurich, 8092 Zurich, Switzerland}
\affiliation{Quantum Center, ETH Zurich, 8083 Zurich, Switzerland}

\author{Nadine Meyer}
\email{nmeyer@ethz.ch}
\affiliation{Nanophotonic Systems Laboratory, Department of Mechanical and Process Engineering, ETH Zurich, 8092 Zurich, Switzerland}
\affiliation{Quantum Center, ETH Zurich, 8083 Zurich, Switzerland}

\author{Romain Quidant}
\affiliation{Nanophotonic Systems Laboratory, Department of Mechanical and Process Engineering, ETH Zurich, 8092 Zurich, Switzerland}
\affiliation{Quantum Center, ETH Zurich, 8083 Zurich, Switzerland}

\date{\today}
\begin{abstract} 
We demonstrate coherent, measurement-free optical feedback control of a levitated nanoparticle, achieving phonon occupations down to a few hundred phonons. Unlike measurement-based feedback, this all-optical scheme preserves the correlations between mechanical motion and the feedback signal. Adjustment of the feedback phase and delay provides precise and tunable control over the system dynamics. The ultimate cooling performance is currently limited by phase noise, which we analyze within a theoretical framework that outlines the constraints and prospects for reaching the motional ground state. Our results establish coherent feedback as a powerful tool for quantum control of levitated systems, extending beyond center-of-mass cooling.
\end{abstract}

\maketitle

Active feedback control is a cornerstone of modern experimental physics widely used to stabilize and manipulate a broad range of systems, ranging from atomic ensemble~\cite{fischer2002feedback,steck2004quantum}, over mechanical oscillators~\cite{Wang2023-fast-feedback-mechanical,  Wilson2015-measurement-control-mechanical,Poggio2007-feedack-cantilever-5mK} to space rockets~\cite{Tewari2011-Control-aircraft-rockets}. In quantum optomechanics, it has been crucial for cooling the motion of mechanical systems to their quantum ground state~\cite{Guo2023-Active-feedback-resonator, Rossi2018-Measurement-based-ground-state} and for state control~\cite{wang2023fast}. 
Such measurement-based feedback extracts information via photodetection which is processed electronically and used to generate corrective forces. While effective, this approach introduces fundamental and technical limitations: measurement backaction disturbs the system~\cite{Purdy2013-pressure-shot-noise-object, Teufel2016-overwhelming-shot-noise}, and electronic noise is reintroduced through the feedback loop~\cite{Poggio2007-feedack-cantilever-5mK}, degrading quantum coherence~\cite{Genes2008-comparing-cold-damping-cavity-assisted}.\\
Coherent feedback offers a fundamentally different approach — one that avoids measurement altogether~\cite{Lloyd2000-coherent-quantum-feedback}. Instead, a light field interacts with the system and is redirected through an optical feedback loop to produce control forces without photo-detection. This measurement-free scheme inherently preserves quantum correlations of light and motion, and avoids the introduction of classical noise through electronics, making it particularly attractive for quantum-limited control. Coherent feedback has already demonstrated promise in the control of light squeezing~\cite{iida2012experimental}, atomic spin systems~\cite{nelson2000experimental, Karg2020-Stron-coupling-membrane-atomic-spin, Schmid2022-coherent-cooling-membrane-atomic-spin}, solid state qubits~\cite{Hirose2016-coherent-feedback-qubit-diamond}, and membranes in optical cavities~\cite{Ernzer2023-coherent-cooling-membrane}, while maintaining quantum coherence.

We explore the application of coherent feedback to a levitated nanoparticle system, a platform that combines excellent environmental isolation~\cite{Dania2024-ultrahigh-quality-levitated} with dynamic control over the potential landscape~\cite{rondin2017direct,hebestreit2018sensing,conangla2020extending,  sun2024tunable, Tomassi2025-inverted-potential}. Recent advances have demonstrated motional ground-state cooling through active, measurement-based feedback~\cite{magrini2021real,tebbenjohanns2021quantum,Kamba2023-revealing-velocity-uncertainty} and via passive optical cavities~\cite{delic2020cooling,piotrowski2022simultaneous,Ranfagni2022-two-dimensional-quantum,dania2024high}. In contrast, here we present the theoretical framework and the experimental realization of measurement-free, coherent feedback to manipulate the center of mass (CoM) motion of an optically trapped nanoparticle. 
Controlled CoM cooling is achieved by adjusting the feedback delay and phase via an optical fiber delay line.

\begin{figure}[h!]
\centering
\includegraphics[width=\linewidth]
{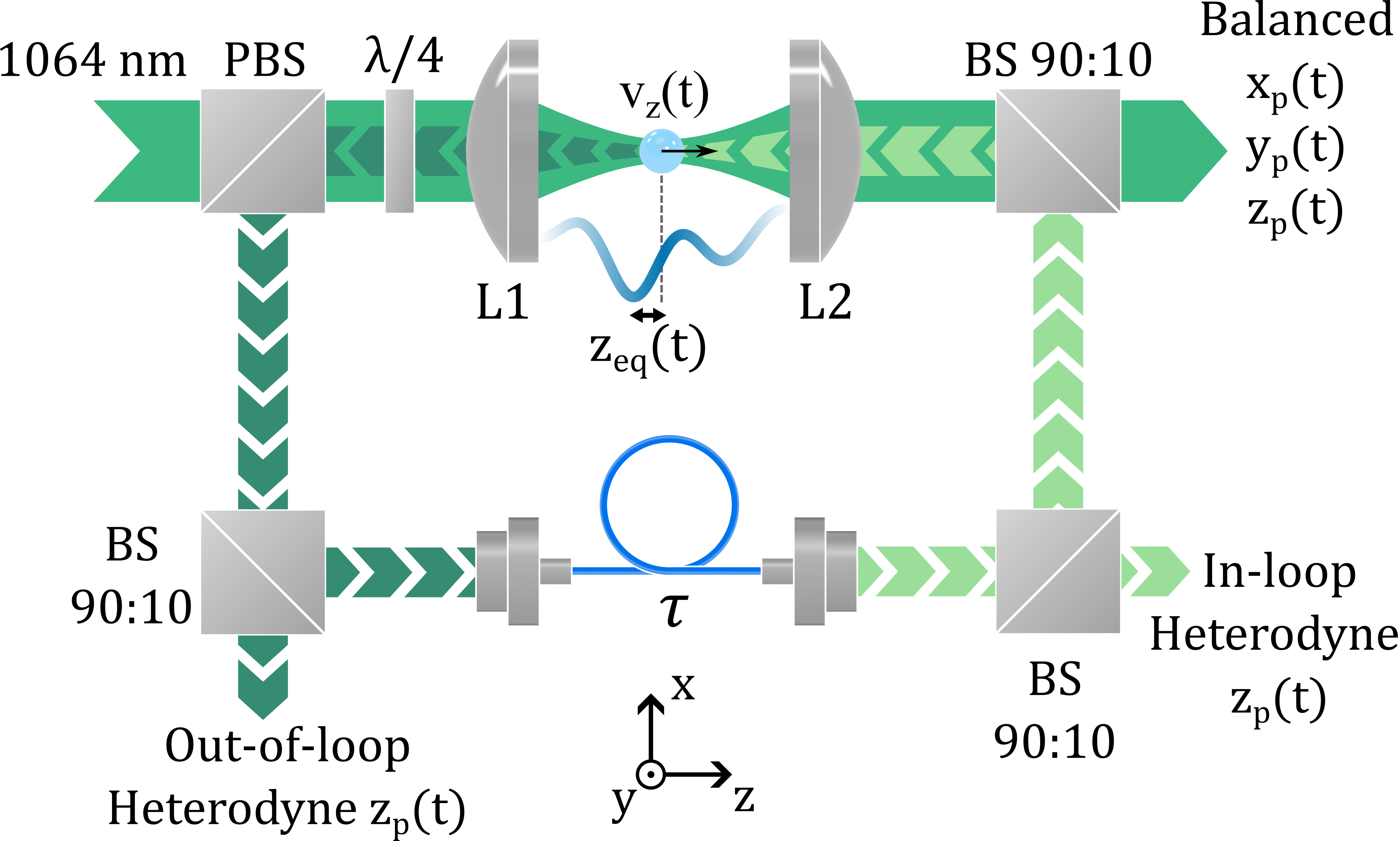}
\caption{
A particle is trapped at the focus of a laser beam (wavelength $\lambda=$\SI{1064}{\nano\meter}, NA=0.77, elliptical polarization). The forward scattered light is collected by lens 2 (L2, NA=0.55) to detect the  particle's motion along $x,y,z$ using balanced detection. The backward scattered light (arrows, dark green) is collected by the trapping lens (L1), separated from the trapping light by a polarizing beam splitter (PBS) and transmitted by an optical fiber of length $L\approx\SI{1.3}{\kilo\meter}$. After a time $\tau$ the scattered light exits the fiber (arrows, light green) and is focused back onto the particle using lens L2.  The interference of counter-propagating scattered and trapping light creates an optical potential (blue curve) with shifted equilibrium $z_\text{eq}(t)$. 
The particle's position along $z_p(t)$ is detected via a heterodyne measurement using part of the back scattered light either in an out-loop measurement (before the delay line) or an in-loop measurement (after the delay line). Additional balanced forward detection of $x_p(t),y_p(t),z_p(t)$ is used to apply electrical cold damping.
}
\label{fig:Setup_simplified}
\end{figure}
\paragraph{General concept}
To assess the relevance of coherent feedback in levitation optomechanics, we demonstrate CoM cooling via linear cold damping and study the limitations of this approach towards ground-state cooling. 
Figure \ref{fig:Setup_simplified} illustrates our coherent optical feedback loop for a dielectric particle trapped in a single-beam optical tweezer (solid beam).

The particle motion, $z_p(t)$, is imprinted on the phase of the scattered light $\phi_p(t)$. 
The back-scattered light (arrows) is partially collected and redirected through an optical feedback loop back onto the particle, counter-propagating to the trapping beam. This so-called feedback beam then interferes with the trapping light, 
creating an interference pattern determined by the relative phase between the counter-propagating beams $\phi(t) = \phi_p(t-\tau) + \phi_0$ where $\phi_0$ is the phase accumulated along the feedback loop. 
Such phase-dependent interference pattern shifts the equilibrium position, $z_\text{eq}(t)$, of the trapping potential, resulting in a position-dependent force on the particle. 
The new equilibrium position, $z_\text{eq}(t)$, is now linked to the position of the particle in the past $z(t-\tau)$, with $\tau$ being the delay accumulated by light through the loop. Assuming the particle behaves as a harmonic oscillator, a delay $\tau=\pi/2\Omega$ generates a velocity-dependent force, $F(t) \propto v_z(t)$, which can be used to cool the particle's motion~\cite{conangla2019optimal,tebbenjohanns2019cold,Vijayan2023-all-optical-scalable-CD}.

\paragraph{Theory}
The equation of motion of a particle trapped by the optical tweezer is

\begin{align}\label{eq:eom}
    \ddot{z}_p(t)+(\Gamma_0+\Gamma_c)\dot{z}_p(t)+\Omega ^2 z_p(t) = \frac{F(t)}{m}
\end{align}
where $z_p(t)$ is the particle's position at time $t$, $m$ its mass and $\Omega$ its mechanical eigenfrequency accounting for additional frequency shifts (see~\cite{SM}). The damping coefficient $\Gamma_0 = \Gamma_m + \Gamma_r+ \Gamma_d$ stems from the surrounding gas $\Gamma_m$~\cite{gieseler2012subkelvin}, the optical trapping field $\Gamma_r$~\cite{jain2016direct} and the electrical linear feedback $\Gamma_d$~\cite{conangla2019optimal}. The damping due to the coherent feedback $\Gamma_c=\beta\Omega\sin(\Omega\tau)$ depends on the linear coefficient $\beta$, relating the delayed particle position and the equilibrium position through $z_{eq}(t)=\beta z_p(t-\tau)$. The parameter $\beta$ depends on the feedback loop efficiency $\eta$, the relative power of the trapping and feedback beams, their polarization and spatial overlap and the numerical aperture of the lenses (see~\cite{SM}).

The stochastic external force $F(t) = \sigma_m \chi_m(t) + \sigma_r \chi_r(t) +  
\sigma_c \chi_c(t)$ relates to residual gas collisions with $\sigma_m=\sqrt{2m k_B T\Gamma_m}$~\cite{gieseler2012subkelvin}, photon recoil with $\sigma_r=\sqrt{(2/5+A^2)\hbar 2\pi P_\text{d}/( c\lambda)}$~\cite{tebbenjohanns2019optimal} 
and phase noise in the feedback beam with $\sigma_c=m\beta\Omega^2\sigma_\phi/B$ (see~\cite{SM}). The geometrical factor $A$ relates to the Gouy phase, $P_d$ to the scattered power, the phase factor $B$ to the phase of the collected light  and the particle's position through $\phi_p(t)=B z_p(t)$, while $\sigma_\phi^2$ relates to optical phase fluctuations. We assume $\chi_i(t)$ to have zero mean and their respective autocorrelation functions to be delta functions with amplitudes $\sigma_i^2$.

From Eq.~\ref{eq:eom} we observe cooling (heating) of the particle motion for $\Gamma_c>0$ ($\Gamma_c<0$). The effect is the largest for maximum interference at $\phi_0=2\ell\pi$, with $\ell\in\mathbb{Z}$, where $\beta$ reaches its extreme values (for other cases see~\cite{SM}).

The effective temperature of the CoM motion is given by the equipartition theorem $T_\text{eff}= m\Omega^2 \langle z_p(t)^2\rangle/ k_B$ where $k_B$ is the Boltzmann constant and  $\langle z_p(t)^2\rangle$ corresponds to the integral of the displacement power spectral density (PSD) $S_{zz}(\omega)$ given by 
\begin{equation}
    S_{zz}(\omega) = \frac{2(\sigma_m^2 + \sigma_r^2 + \sigma_c^2)}{m^2((\Omega^2 - \omega^2)^2 + \omega^2 (\Gamma_0+\Gamma_c)^2)}
\end{equation}

The effective temperature is given by 
\begin{equation}\label{eq:Teff2}
    T_\text{eff}= \frac{m\Omega^2}{2k_B} \left( \frac{\sigma_m^2 + \sigma_r^2}{m^2 \Omega^2(\Gamma_0 + \Gamma_c)} + \frac{\sigma_\phi^2\Omega^2}{B^2}\frac{\beta^2}{\Gamma_0 + \Gamma_c} \right).
\end{equation}

Hereafter, we need to distinguish two cooling scenarios: For weak cooling (small $\beta$), the second term in Eq.~\ref{eq:Teff2} is negligible, in contrast to strong cooling (large $\beta$) where both terms contribute similarly. For weak cooling with small $\beta$, Eq.~\ref{eq:Teff2} takes the well-known expression~\cite{gieseler2012subkelvin}

\begin{equation}\label{eq:Teff_weak}
       T_\text{eff} =  \frac{T_0 \Gamma_0}{\Gamma_0 + \Gamma_c}  = \frac{\sigma_m^2 + \sigma_r^2}{2k_B m (\Gamma_0 +\beta\Omega\sin(\Omega\tau))} 
\end{equation}
where $T_0$ is the temperature in the absence of coherent feedback ($\beta =0$ or $\Omega\tau= \pi)$. 
Eq.~\ref{eq:Teff_weak} suggests that the temperature decreases for increasing $\beta$ and is optimized for a delay $\tau= \pi/(2\Omega)$ corresponding to a quarter of the oscillation period.

In contrast, for strong cooling with large $\beta$, the noise in the feedback force correlates with the particle motion, since the displacement of the equilibrium position $z_\text{eq}$ becomes dominated by phase noise. This effect is similar to measurement-based cold damping~\cite{conangla2019optimal}. 

Since the two terms in Eq.~\ref{eq:Teff2} scale differently with $\beta$, there exists an optimal feedback strength $\beta_\text{opt}= \sqrt{\sigma_m^2 + \sigma_r^2} B/(m\Omega^2 \sigma_\phi)$
for which a minimum temperature 
\begin{equation}\label{eq:Tmin}
    T_\text{eff,min} = \frac{\sigma_\phi\Omega\sqrt{\sigma_m^2+\sigma_r^2}}{k_B B \sin(\Omega\tau)}
\end{equation}is achieved where we assumed $\Gamma_0 \ll\Gamma_c$. 
From Eq.~\ref{eq:Tmin}, we observe an increase in the minimal temperature with increasing noise $\sigma_\phi,\sigma_m$ and $\sigma_r$. 
Increasing the parameter beyond $\beta_\text{opt}$ ultimately heats the particle and the detected PSD shows signatures of noise squashing (see~\cite{SM}). 

\paragraph{Experimental implementation}
To study the effect of coherent feedback, a charged silica nanoparticle of radius $R \approx \SI{97}{\nano\meter}$ and mass $m \approx \SI{7.07}{\femto\gram}$ is trapped by an optical tweezer along the optical axis $z$ (large green beam) of elliptically polarized light at wavelength $\lambda = \SI{1064}{\nano\meter}$, focused by lens 1 of numerical aperture $\text{NA}_1=0.77$ and the geometrical factor due to the Gouy-phase $A_1$= 0.83 as depicted in Fig.~\ref{fig:Setup_simplified}. At the power $P_t\approx\SI{300}{\milli\watt}$, the mechanical resonance frequencies of the particle are $(\Omega_x,\Omega_y,\Omega )/2\pi\approx(187, 162, 47)$ kHz. The light scattered forward by the particle is collected by lens 2 with numerical aperture $\text{NA}_2=0.55$ and used to detect the position of the particle in a balanced detection scheme~\cite{gieseler2012subkelvin}. This position measurement is used to stabilize the particle in 3D to a base temperature $T_0\ll \SI{300}{\kelvin}$ by means of linear electrical cold damping ($\Gamma_d>0$) ~\cite{conangla2019optimal,Tomassi2025-inverted-potential}. 

The light scattered backwards by the particle (dark green arrows) is collected using lens 1 and separated from the trapping beam by a polarizing beam splitter (PBS). This feedback beam passes through a single-mode fiber of length $L\approx \SI{1.3}{\kilo\meter}$ and refractive index $n_\text{eff}\approx1.46$, causing a delay time $\tau = L n_\text{eff}/c\approx \SI{6.34}{\micro\second}$. The collected light (light green arrows) is focused on the particle by lens 2, closing the feedback loop with a total efficiency $\eta$. The accumulated phase $\phi_0$ along the feedback loop is controlled by changing the beam path length with a mirror placed on a piezo-electric element. For the coherent feedback, we use part of the feedback light in an additional heterodyne measurement~\cite{magrini2021real,tebbenjohanns2021quantum} to detect the particle motion along $z$ (see Fig.~\ref{fig:Setup_simplified}). The heterodyne detector is placed either in front of the delay line (out-of loop measurement, OL) or after (in-loop measurement, IL).  
All data presented in this manuscript is acquired in backwards heterodyne detection.

\begin{figure}
\centering
\includegraphics[width=\linewidth]
{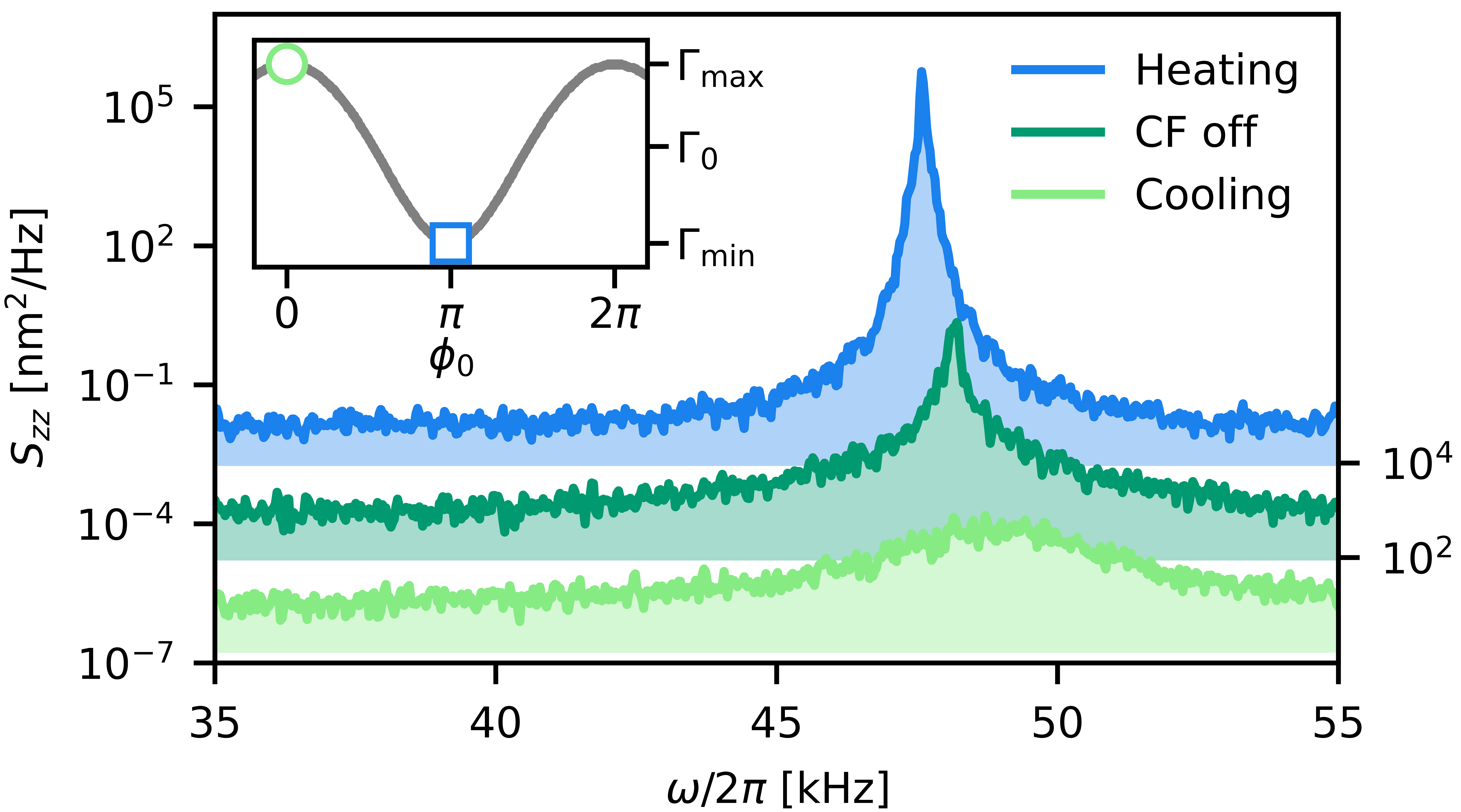}
\caption{PSD of particle displacement with coherent feedback at $p=1\times10^{-4}\SI{}{\milli\bar}$.  Depending on the phase $\phi_0$, the coherent feedback cools (light green, circle) or heats (blue, square) the particle motion in comparison to the absence of feedback light (dark green). The individual particle displacements are offset to increase visibility. Inset: For $\phi_0 = 2n\pi$, the damping $\Gamma_c>0$ is maximal (light green circle), while for $\phi_0 = (2n+1)\pi$, $\Gamma_c<0$ is minimal (blue square). 
}
\label{fig:Heating_and_cooling}
\end{figure}

To ensure the linear cooling regime described by Eq.~\ref{eq:Teff_weak}, the feedback beam is optically attenuated, reducing $|\beta|$. For a delay $\tau\Omega \approx \pi/2$, a velocity dependent force heats ($\Gamma_c <0$) or cools  ($\Gamma_c >0$) the particle motion, depending on $\phi_0$ as depicted in Fig.~\ref{fig:Heating_and_cooling}. In comparison to the motion in absence of coherent feedback  
(dark green), the coherent feedback cools for $\phi_0=0$ with $\Gamma_c>0$ (light green, circle) and heats for $\phi_0 = \pi$ with $\Gamma_c<0$ (blue, square), represented by the decreasing or increasing area of the PSDs.
The observed damping rate reaches $(\Gamma_0+\Gamma_c)/2\pi\approx\SI{1.92}{\kilo\hertz}$ at maximum cooling, corresponding to $\beta\approx4.10\times10^{-2}$, while the omnipresent damping rate equals $\Gamma_0/2\pi\approx\SI{78}{\hertz}$ (see~\cite{SM}). 

\begin{figure}
\centering
\includegraphics[width=\linewidth]
{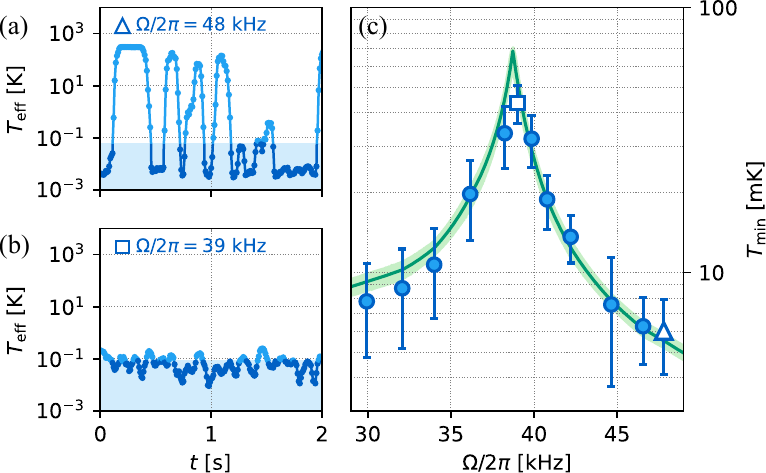}
\caption{\textbf{(a)-(b)} CoM temperature without phase lock  (free-running $\phi_0$) leading to random cooling and heating in time. \textbf{(a)} At the resonance frequency $\Omega/(2\pi) = \SI{48}{\kilo\hertz}$  with $\Omega\tau'>\pi$, strong coherent feedback manifests itself in high maximum and low minimum temperatures due to $\text{max}(|\Gamma_c|)\gg\Gamma_0$. \textbf{(b)} In contrast, the coherent feedback is ineffective for $\Omega/(2\pi)=\SI{39}{\kilo\hertz}$  with $\Omega\tau'\approx  \pi$ leading to a reduced variation in $T_\text{eff}$ due to $|\Gamma_c|\ll\Gamma_0$.
\textbf{(c)} Minimum achieved temperature of the CoM $T_\text{min}$ for different resonance frequencies $\Omega$. For mismatched delays at $\Omega/(2\pi)\approx\SI{39}{\kilo\hertz}$ the cooling is inefficient ($\Omega\tau'\approx\pi$). For smaller and larger $\Omega$ the delay is more optimized and the coherent feedback cools efficiently to lower $T_\text{min}$. Each point is the average of 5 measurements of $T_\text{min}$, and the error bars represent their standard deviation. $T_\text{min}$ is the median of the blue shaded region in \textbf{(a)-(b)} corresponding to $T_\text{eff} < T_0(\Omega) +2 \Sigma(T_0(\Omega))$. The data is fitted to Eq.~\ref{eq:fit_Fig3} (green line) taking the fitting errors into account (green shaded area).}
\label{fig:T_vs_fz}
\end{figure}

To demonstrate the effect of the resonance frequency $\Omega$ on the cooling efficiency, we measure the temperature $T_\text{eff}$ for varying $\Omega$ which corresponds to different $\Gamma_c\propto \Omega \sin(\Omega\tau)$. The resonance frequency $\Omega$ is varied by changing the optical tweezer power such that $\beta$ remains constant (see~\cite{SM}). We double the delay to $\tau'=2\tau$ by placing a retro-reflector at the end of the delay line followed by a circulator to route the feedback beam back to the particle (see Fig.~S6). Note that the phase $\phi_0$ remains now uncontrolled. Fig~\ref{fig:T_vs_fz}(a) and (b) display $T_\text{eff}(\Omega )$ in time for two  different $\Omega\tau$. We estimate $T_\text{eff}(\Omega )$ from the area of the PSD of the particle displacement 
(see~\cite{SM}).
The fluctuations of $\phi_0$ govern if the particle motion is cooled or heated.  For a delay $\Omega\tau'>\pi$ at $\Omega/(2\pi)=\SI{48}{\kilo\hertz}$, as in Fig.~\ref{fig:T_vs_fz}(a), the lowest temperature and pronounced temperature variations are observed, suggesting strong coherent feedback ($\Gamma_c \gg \Gamma_0$). In contrast, for $\Omega\tau' \approx \pi$ at $\Omega/(2\pi)=\SI{39}{\kilo\hertz}$, as in Fig.~\ref{fig:T_vs_fz}(b), the particle motion remains mainly unaffected for all time as expected from a weak feedback strength ($\Gamma_c\approx  0$). 

Fig.~\ref{fig:T_vs_fz}(c) depicts $T_\text{min}(\Omega )$ in dependence of the resonance frequency $\Omega$, where each data point corresponds to the median of all instances of $T_\text{eff}(\Omega )<T_0(\Omega) +2 \Sigma(T_0(\Omega))$ where $\Sigma$ is the standard deviation (blue shaded region). The green solid curve represents a fit to 
\begin{equation}\label{eq:fit_Fig3}
    \frac{1}{T_\text{eff}(\Omega )}-\frac{1}{T_0(\Omega )}=\frac{\beta}{\Gamma_0(\Omega)T_0(\Omega)}\Omega \sin{\Omega \tau'}
\end{equation} 
where the fitting parameters are $\beta/(\Gamma_0 T_0)$ and $\tau'$. The fitted value $\tau'=12.9 \pm \SI{0.04}{\micro\second}$ is in good agreement with the expected value $\tau' \approx \SI{12.7}{\micro\second}$. From the fit we extract $\beta\approx(6.18 \pm 0.4)\times10^{-4}$, corresponding to $\Gamma_c/2\pi\approx\SI{12}{\hertz}$ for $\Omega\tau' \approx 3\pi/4$ (Fig.~\ref{fig:T_vs_fz}(a))  and $\Gamma_c/2\pi\approx\SI{0.55}{\hertz}$ for $\Omega\tau' \approx \pi/2$ (Fig.~\ref{fig:T_vs_fz}(b)), where the feedback beam has been optically attenuated in comparison to  Fig~\ref{fig:Heating_and_cooling}. 

\begin{figure}
\centering
\includegraphics[width=\linewidth]
{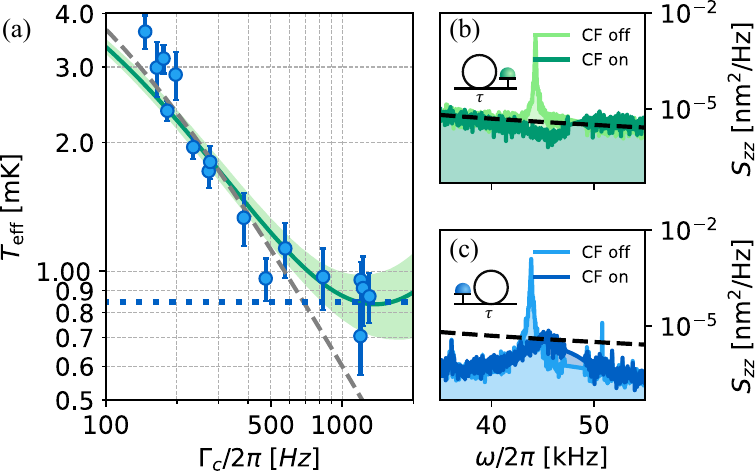}
\caption{
\textbf{(a)}  $T_\text{eff}$ versus $\Gamma_c$ by varying the feedback power. The minimum temperature achieved is $T_\text{eff}\approx705 \pm 133$\SI{}{\micro\kelvin}, corresponding to $n =  k_B T_\text{eff} /(\hbar\Omega) = 344 \pm 55$ phonons. Data and errorbars represent the mean and standard deviation of 5 measurements. The data is fitted to Eq.~\ref{eq:Teff2} (green solid line)  with the corresponding fitting error (light green shaded region).
For weak cooling below the optimal $\Gamma_c$, the hottest eight points are fitted to Eq.~\ref{eq:Teff_weak} (dashed line) neglecting the phase noise heating at high values of $\Gamma_c$. Eq.~\ref{eq:Tmin} (dotted line) predicts the minimum reachable temperature. \textbf{(b)} In-loop PSD of the particle displacement detected after the delay line. Without coherent feedback (CF) for $\Gamma_c =0$ (light green), the particle motion is clearly visible, while with CF for $\Gamma_c > 0$ (dark green), noise squashing indicates that the feedback is dominated by phase noise.  The displacement noise due to the phase noise (black dashed line) matches the data (see Eq.~S37 in~\cite{SM}). 
\textbf{(c)} Out-of-loop PSD of the particle displacement in front of the delay line. When the CF is on ($\Gamma_c > 0$ (dark blue), the particle amplitude is reduced in comparison to $\Gamma_c =0$ (light blue) but clearly visible. The peak value of the PSD matches the displacement noise observed in the in-loop detector. 
}
\label{fig:T_vs_kd}
\end{figure}

To estimate the minimum achievable temperature $T_\text{eff,min}$ in our system, the feedback phase is stabilized at $\phi_0\approx0$ and $\Gamma_c$ is adjusted by attenuating the power of the feedback beam, thereby emulating a variable feedback loop efficiency $\eta$, with $\Gamma_c \propto \sqrt{\eta}$ (see~\cite{SM}). Additionally, we maximize $\Omega$ to benefit from lower phase noise. Fig.~\ref{fig:T_vs_kd}(a) displays the linear decrease of $T_\text{eff}(\Omega )$ for increasing but small $\Gamma_c$ (blue circles) as expected from Eq.~\ref{eq:Teff_weak}. For large enough $\Gamma_c$, $T_\text{eff}(\Omega )$  starts leveling off despite increasing feedback gain. This effect is due to the fact that phase noise starts dominating over the phase imprinted by the particle motion $\phi_p(t)$, heating the particle motion as predicted by Eq.~\ref{eq:Teff2}. At $p\approx 3\times 10^{-7}\SI{}{\milli\bar}$, we measure a minimum temperature  $T_\text{eff,min}\approx 705 \pm \SI{133}{\micro\kelvin}$ at $\Gamma_c \approx 2\pi\times \SI{250}{\hertz}$, corresponding to $n = k_B T_\text{eff} /(\hbar\Omega) =344 \pm 55$ phonons. Note that $\Gamma_c\simeq 10 \Gamma_d$ and  therefore the particle motion is predominantly cooled by coherent feedback. For small $\Gamma_c$ the high temperature data points fulfilling $T_\text{eff}(\Omega )>\SI{1.5}{\milli\K}$ are fitted to Eq.~\ref{eq:Teff_weak} (dashed gray line), suggesting a monotonic decrease in $T_\text{eff}(\Omega )$. In contrast, the fit to  Eq.~\ref{eq:Teff2} (green solid curve) takes all dominating noise sources into account and captures the existence of a minimum temperature $T_\text{min}$. According to Eq.~\ref{eq:Tmin}, the minimum temperature (blue dotted line) equals $T_\text{min} = \SI{847}{\micro\kelvin}$, which is in good agreement with the experiment. 

The main limitation of the minimal effective temperature $T_\text{eff,min}$ is the phase noise $\sigma_\phi$ introduced by the delay line. 
We compare the PSD of the particle's displacement at the IL (Fig.~\ref{fig:T_vs_kd}(b)) and OL detector (Fig.~\ref{fig:T_vs_kd}(c)) with (dark points)  and without coherent feedback (light points). Here, $\Gamma_c$ equals the largest value used in Fig.~\ref{fig:T_vs_kd}(a). The IL detector shows noise squashing, a dip below the noise floor (dashed line), due to increased phase noise once coherent feedback is applied (dark green). 
In contrast, the signal-to-noise ratio remains well above one at the OL detector even when the coherent feedback is applied (dark blue). Note that the particle is always exposed to the accumulated phase noise, since the delay line is required for cooling. 
The displacement noise at $\Omega$ for $\beta_\text{opt}$ is extracted from a fit to the noise floor of the IL detector to $S_{zz,\text{IL}}(\Omega) \approx 2\sigma_{\phi}^2/B^2= 4\times10^{-24}\text{m}^2/\text{Hz}$ (dashed line), corresponding to the phase noise contribution (see Fig.~\ref{fig:T_vs_kd}(b)).

\paragraph{Discussion} 
With the aim of ground state cooling with coherent feedback, the minimum achievable temperature  described by Eq.~\ref{eq:Tmin} shows that lowering phase noise $\sigma_\phi$, residual gas collisions $\sigma_m$, and photon recoil $\sigma_r$ is beneficial. 
The influence of $\Omega$ on $T_{\text{eff,min}}$ is more complex. An increase in $\Omega$ results in a reduction of the required delay line length $L$. However, their effects on the minimum effective temperature tend to compensate, since $T_{\text{eff,min}} \propto \Omega / L$, under the assumption that $\sigma_\phi \propto L$. On the other hand, adopting the common assumption that $\sigma_\phi \propto 1/\Omega$ leads to an overall decrease in the minimum effective temperature.
In a backaction-limited experiment, smaller particles are also advantageous due to $T_\text{min} \propto R^3$, and enhanced collection efficiency further contributes positively.
In current levitation experiments, phase noise~\cite{kamba2021recoil, dania2024high} and residual gas pressure~\cite{magrini2021real, Dania2024-ultrahigh-quality-levitated, Lindner2024-hollow-core-ultra-high-vacuum} have shown to be subject of significant improvements. Assuming a phase noise reduction by 30dB~\cite{chao2025robust,parniak2021high}, and a $2.5\times$ smaller particle promises at  $p=1\times 10^{-8}$ mbar~\cite{magrini2021real,Dania2024-ultrahigh-quality-levitated} a minimum phonon occupation of 0.9 phonons, reaching ground state.

\paragraph{Outlook and conclusion}
In summary, we have demonstrated the impact of measurement-free coherent feedback on the motion of a levitated particle. Our implementation employs an all-optical feedback loop that can preserve the correlations between the mechanical motion and the feedback signal. By tuning the feedback phase and its delay, we achieved full control over the system dynamics. Furthermore, we presented a theoretical framework that outlines the fundamental limitations to ground-state cooling in this configuration.
Levitated nanoparticles are inherently low-frequency oscillators, which necessitates a longer feedback delay $\tau$ compared to other optomechanical systems~\cite{Ernzer2023-coherent-cooling-membrane}. This extended delay introduces increased phase noise, which scales with the delay line length. In many practical applications, phase noise may also be frequency dependent, often becoming more significant at lower frequencies. As a result, levitation may initially appear less suitable for coherent feedback cooling. However, levitation in high-frequency regimes — such as librational degrees of freedom~\cite{dania2024high,pontin2023simultaneous,gao2024feedback,bang2020five}, levitation in standing-wave optical traps~\cite{kamba2021recoil}, or using high-refractive-index materials~\cite{lepeshov2023levitated,afridi2025controlling,lee2025inverse} — offers more promising prospects.
Additionally, unique three-dimensional cooling of levitated particles could be enabled through multidirectional detection schemes~\cite{Dinter2024-3d-displacement-sensing-mode-decomposition}, such as those implemented on integrated platforms~\cite{Melo2024-levitation-on-chip}.
Importantly, coherent feedback has applications beyond cooling. It can be employed to engineer a variety of system interactions, including dissipative and nonreciprocal dynamics~\cite{Karg2019-remote-hamiltonian-mediated-light,metelmann2015nonreciprocal}, optomechanics with Fanoresonators~\cite{du2025coherent} and motional entanglement~\cite{Lloyd2000-coherent-quantum-feedback,li2017enhanced,Zambon2025-remote-entanglement-particles}. Some of these approaches may circumvent the need for long delay lines and thus avoid the associated phase noise, making coherent feedback a powerful tool for advancing quantum control of levitated particles.

\paragraph{Aknowledgments}
We acknowledge valuable discussions with P. Treutlein and G.-L. Schmidt, and the QXtreme consortium. This research was supported by the European Research Council (ERC) under the grant Agreement No. [951234] (Q-Xtreme ERC-2020-SyG).


\bibliography{references}

\begin{thebibliography}{56}%
\makeatletter
\providecommand \@ifxundefined [1]{%
 \@ifx{#1\undefined}
}%
\providecommand \@ifnum [1]{%
 \ifnum #1\expandafter \@firstoftwo
 \else \expandafter \@secondoftwo
 \fi
}%
\providecommand \@ifx [1]{%
 \ifx #1\expandafter \@firstoftwo
 \else \expandafter \@secondoftwo
 \fi
}%
\providecommand \natexlab [1]{#1}%
\providecommand \enquote  [1]{``#1''}%
\providecommand \bibnamefont  [1]{#1}%
\providecommand \bibfnamefont [1]{#1}%
\providecommand \citenamefont [1]{#1}%
\providecommand \href@noop [0]{\@secondoftwo}%
\providecommand \href [0]{\begingroup \@sanitize@url \@href}%
\providecommand \@href[1]{\@@startlink{#1}\@@href}%
\providecommand \@@href[1]{\endgroup#1\@@endlink}%
\providecommand \@sanitize@url [0]{\catcode `\\12\catcode `\$12\catcode
  `\&12\catcode `\#12\catcode `\^12\catcode `\_12\catcode `\%12\relax}%
\providecommand \@@startlink[1]{}%
\providecommand \@@endlink[0]{}%
\providecommand \url  [0]{\begingroup\@sanitize@url \@url }%
\providecommand \@url [1]{\endgroup\@href {#1}{\urlprefix }}%
\providecommand \urlprefix  [0]{URL }%
\providecommand \Eprint [0]{\href }%
\providecommand \doibase [0]{https://doi.org/}%
\providecommand \selectlanguage [0]{\@gobble}%
\providecommand \bibinfo  [0]{\@secondoftwo}%
\providecommand \bibfield  [0]{\@secondoftwo}%
\providecommand \translation [1]{[#1]}%
\providecommand \BibitemOpen [0]{}%
\providecommand \bibitemStop [0]{}%
\providecommand \bibitemNoStop [0]{.\EOS\space}%
\providecommand \EOS [0]{\spacefactor3000\relax}%
\providecommand \BibitemShut  [1]{\csname bibitem#1\endcsname}%
\let\auto@bib@innerbib\@empty
\bibitem [{\citenamefont {Fischer}\ \emph {et~al.}(2002)\citenamefont
  {Fischer}, \citenamefont {Maunz}, \citenamefont {Pinkse}, \citenamefont
  {Puppe},\ and\ \citenamefont {Rempe}}]{fischer2002feedback}%
  \BibitemOpen
  \bibfield  {author} {\bibinfo {author} {\bibfnamefont {T.}~\bibnamefont
  {Fischer}}, \bibinfo {author} {\bibfnamefont {P.}~\bibnamefont {Maunz}},
  \bibinfo {author} {\bibfnamefont {P.}~\bibnamefont {Pinkse}}, \bibinfo
  {author} {\bibfnamefont {T.}~\bibnamefont {Puppe}},\ and\ \bibinfo {author}
  {\bibfnamefont {G.}~\bibnamefont {Rempe}},\ }\bibfield  {title} {\bibinfo
  {title} {Feedback on the motion of a single atom in an optical cavity},\
  }\href {https://journals.aps.org/prl/abstract/10.1103/PhysRevLett.88.163002}
  {\bibfield  {journal} {\bibinfo  {journal} {Phys. Rev. Lett.}\ }\textbf
  {\bibinfo {volume} {88}},\ \bibinfo {pages} {163002} (\bibinfo {year}
  {2002})}\BibitemShut {NoStop}%
\bibitem [{\citenamefont {Steck}\ \emph {et~al.}(2004)\citenamefont {Steck},
  \citenamefont {Jacobs}, \citenamefont {Mabuchi}, \citenamefont
  {Bhattacharya},\ and\ \citenamefont {Habib}}]{steck2004quantum}%
  \BibitemOpen
  \bibfield  {author} {\bibinfo {author} {\bibfnamefont {D.~A.}\ \bibnamefont
  {Steck}}, \bibinfo {author} {\bibfnamefont {K.}~\bibnamefont {Jacobs}},
  \bibinfo {author} {\bibfnamefont {H.}~\bibnamefont {Mabuchi}}, \bibinfo
  {author} {\bibfnamefont {T.}~\bibnamefont {Bhattacharya}},\ and\ \bibinfo
  {author} {\bibfnamefont {S.}~\bibnamefont {Habib}},\ }\bibfield  {title}
  {\bibinfo {title} {Quantum feedback control of atomic motion in an optical
  cavity},\ }\href
  {https://journals.aps.org/prl/abstract/10.1103/PhysRevLett.92.223004}
  {\bibfield  {journal} {\bibinfo  {journal} {Phys. Rev. Lett.}\ }\textbf
  {\bibinfo {volume} {92}},\ \bibinfo {pages} {223004} (\bibinfo {year}
  {2004})}\BibitemShut {NoStop}%
\bibitem [{\citenamefont {Wang}\ \emph
  {et~al.}(2023{\natexlab{a}})\citenamefont {Wang}, \citenamefont {Banniard},
  \citenamefont {de~L\'epinay},\ and\ \citenamefont
  {Sillanp\"a\"a}}]{Wang2023-fast-feedback-mechanical}%
  \BibitemOpen
  \bibfield  {author} {\bibinfo {author} {\bibfnamefont {C.}~\bibnamefont
  {Wang}}, \bibinfo {author} {\bibfnamefont {L.}~\bibnamefont {Banniard}},
  \bibinfo {author} {\bibfnamefont {L.~M.}\ \bibnamefont {de~L\'epinay}},\ and\
  \bibinfo {author} {\bibfnamefont {M.~A.}\ \bibnamefont {Sillanp\"a\"a}},\
  }\bibfield  {title} {\bibinfo {title} {Fast feedback control of mechanical
  motion using circuit optomechanics},\ }\href
  {https://doi.org/10.1103/PhysRevApplied.19.054091} {\bibfield  {journal}
  {\bibinfo  {journal} {Phys. Rev. Appl.}\ }\textbf {\bibinfo {volume} {19}},\
  \bibinfo {pages} {054091} (\bibinfo {year} {2023}{\natexlab{a}})}\BibitemShut
  {NoStop}%
\bibitem [{\citenamefont {Wilson}\ \emph {et~al.}(2015)\citenamefont {Wilson},
  \citenamefont {Sudhir}, \citenamefont {Piro}, \citenamefont {Schilling},
  \citenamefont {Ghadimi},\ and\ \citenamefont
  {Kippenberg}}]{Wilson2015-measurement-control-mechanical}%
  \BibitemOpen
  \bibfield  {author} {\bibinfo {author} {\bibfnamefont {D.~J.}\ \bibnamefont
  {Wilson}}, \bibinfo {author} {\bibfnamefont {V.}~\bibnamefont {Sudhir}},
  \bibinfo {author} {\bibfnamefont {N.}~\bibnamefont {Piro}}, \bibinfo {author}
  {\bibfnamefont {R.}~\bibnamefont {Schilling}}, \bibinfo {author}
  {\bibfnamefont {A.}~\bibnamefont {Ghadimi}},\ and\ \bibinfo {author}
  {\bibfnamefont {T.~J.}\ \bibnamefont {Kippenberg}},\ }\bibfield  {title}
  {\bibinfo {title} {Measurement-based control of a mechanical oscillator at
  its thermal decoherence rate},\ }\href
  {https://www.nature.com/articles/nature14672} {\bibfield  {journal} {\bibinfo
   {journal} {Nature}\ }\textbf {\bibinfo {volume} {524}},\ \bibinfo {pages}
  {325} (\bibinfo {year} {2015})}\BibitemShut {NoStop}%
\bibitem [{\citenamefont {Poggio}\ \emph {et~al.}(2007)\citenamefont {Poggio},
  \citenamefont {Degen}, \citenamefont {Mamin},\ and\ \citenamefont
  {Rugar}}]{Poggio2007-feedack-cantilever-5mK}%
  \BibitemOpen
  \bibfield  {author} {\bibinfo {author} {\bibfnamefont {M.}~\bibnamefont
  {Poggio}}, \bibinfo {author} {\bibfnamefont {C.~L.}\ \bibnamefont {Degen}},
  \bibinfo {author} {\bibfnamefont {H.~J.}\ \bibnamefont {Mamin}},\ and\
  \bibinfo {author} {\bibfnamefont {D.}~\bibnamefont {Rugar}},\ }\bibfield
  {title} {\bibinfo {title} {Feedback cooling of a cantilever's fundamental
  mode below 5 mk},\ }\href {https://doi.org/10.1103/PhysRevLett.99.017201}
  {\bibfield  {journal} {\bibinfo  {journal} {Phys. Rev. Lett.}\ }\textbf
  {\bibinfo {volume} {99}},\ \bibinfo {pages} {017201} (\bibinfo {year}
  {2007})}\BibitemShut {NoStop}%
\bibitem [{\citenamefont {Tewari}(2011)}]{Tewari2011-Control-aircraft-rockets}%
  \BibitemOpen
  \bibfield  {author} {\bibinfo {author} {\bibfnamefont {A.}~\bibnamefont
  {Tewari}},\ }\href
  {https://onlinelibrary.wiley.com/doi/book/10.1002/9781119971191} {\emph
  {\bibinfo {title} {Advanced control of aircraft, spacecraft and rockets}}},\
  Aerospace Series\ (\bibinfo  {publisher} {Wiley-Blackwell},\ \bibinfo
  {address} {Hoboken, NJ},\ \bibinfo {year} {2011})\BibitemShut {NoStop}%
\bibitem [{\citenamefont {Guo}\ \emph {et~al.}(2023)\citenamefont {Guo},
  \citenamefont {Chang}, \citenamefont {Yao},\ and\ \citenamefont
  {Gr{\"o}blacher}}]{Guo2023-Active-feedback-resonator}%
  \BibitemOpen
  \bibfield  {author} {\bibinfo {author} {\bibfnamefont {J.}~\bibnamefont
  {Guo}}, \bibinfo {author} {\bibfnamefont {J.}~\bibnamefont {Chang}}, \bibinfo
  {author} {\bibfnamefont {X.}~\bibnamefont {Yao}},\ and\ \bibinfo {author}
  {\bibfnamefont {S.}~\bibnamefont {Gr{\"o}blacher}},\ }\bibfield  {title}
  {\bibinfo {title} {Active-feedback quantum control of an integrated
  low-frequency mechanical resonator},\ }\href
  {https://www.nature.com/articles/s41467-023-40442-3} {\bibfield  {journal}
  {\bibinfo  {journal} {Nat. Commun.}\ }\textbf {\bibinfo {volume} {14}},\
  \bibinfo {pages} {4721} (\bibinfo {year} {2023})}\BibitemShut {NoStop}%
\bibitem [{\citenamefont {Rossi}\ \emph {et~al.}(2018)\citenamefont {Rossi},
  \citenamefont {Mason}, \citenamefont {Chen}, \citenamefont {Tsaturyan},\ and\
  \citenamefont {Schliesser}}]{Rossi2018-Measurement-based-ground-state}%
  \BibitemOpen
  \bibfield  {author} {\bibinfo {author} {\bibfnamefont {M.}~\bibnamefont
  {Rossi}}, \bibinfo {author} {\bibfnamefont {D.}~\bibnamefont {Mason}},
  \bibinfo {author} {\bibfnamefont {J.}~\bibnamefont {Chen}}, \bibinfo {author}
  {\bibfnamefont {Y.}~\bibnamefont {Tsaturyan}},\ and\ \bibinfo {author}
  {\bibfnamefont {A.}~\bibnamefont {Schliesser}},\ }\bibfield  {title}
  {\bibinfo {title} {Measurement-based quantum control of mechanical motion},\
  }\href {https://www.nature.com/articles/s41586-018-0643-8} {\bibfield
  {journal} {\bibinfo  {journal} {Nature}\ }\textbf {\bibinfo {volume} {563}},\
  \bibinfo {pages} {53} (\bibinfo {year} {2018})}\BibitemShut {NoStop}%
\bibitem [{\citenamefont {Wang}\ \emph
  {et~al.}(2023{\natexlab{b}})\citenamefont {Wang}, \citenamefont {Banniard},
  \citenamefont {de~L{\'e}pinay},\ and\ \citenamefont
  {Sillanp{\"a}{\"a}}}]{wang2023fast}%
  \BibitemOpen
  \bibfield  {author} {\bibinfo {author} {\bibfnamefont {C.}~\bibnamefont
  {Wang}}, \bibinfo {author} {\bibfnamefont {L.}~\bibnamefont {Banniard}},
  \bibinfo {author} {\bibfnamefont {L.~M.}\ \bibnamefont {de~L{\'e}pinay}},\
  and\ \bibinfo {author} {\bibfnamefont {M.~A.}\ \bibnamefont
  {Sillanp{\"a}{\"a}}},\ }\bibfield  {title} {\bibinfo {title} {Fast feedback
  control of mechanical motion using circuit optomechanics},\ }\href
  {https://journals.aps.org/prapplied/abstract/10.1103/PhysRevApplied.19.054091}
  {\bibfield  {journal} {\bibinfo  {journal} {Phys. Rev. Appl.}\ }\textbf
  {\bibinfo {volume} {19}},\ \bibinfo {pages} {054091} (\bibinfo {year}
  {2023}{\natexlab{b}})}\BibitemShut {NoStop}%
\bibitem [{\citenamefont {Purdy}\ \emph {et~al.}(2013)\citenamefont {Purdy},
  \citenamefont {Peterson},\ and\ \citenamefont
  {Regal}}]{Purdy2013-pressure-shot-noise-object}%
  \BibitemOpen
  \bibfield  {author} {\bibinfo {author} {\bibfnamefont {T.~P.}\ \bibnamefont
  {Purdy}}, \bibinfo {author} {\bibfnamefont {R.~W.}\ \bibnamefont
  {Peterson}},\ and\ \bibinfo {author} {\bibfnamefont {C.~A.}\ \bibnamefont
  {Regal}},\ }\bibfield  {title} {\bibinfo {title} {Observation of radiation
  pressure shot noise on a macroscopic object},\ }\href
  {https://www.science.org/doi/10.1126/science.1231282} {\bibfield  {journal}
  {\bibinfo  {journal} {Science}\ }\textbf {\bibinfo {volume} {339}},\ \bibinfo
  {pages} {801} (\bibinfo {year} {2013})}\BibitemShut {NoStop}%
\bibitem [{\citenamefont {Teufel}\ \emph {et~al.}(2016)\citenamefont {Teufel},
  \citenamefont {Lecocq},\ and\ \citenamefont
  {Simmonds}}]{Teufel2016-overwhelming-shot-noise}%
  \BibitemOpen
  \bibfield  {author} {\bibinfo {author} {\bibfnamefont {J.~D.}\ \bibnamefont
  {Teufel}}, \bibinfo {author} {\bibfnamefont {F.}~\bibnamefont {Lecocq}},\
  and\ \bibinfo {author} {\bibfnamefont {R.~W.}\ \bibnamefont {Simmonds}},\
  }\bibfield  {title} {\bibinfo {title} {Overwhelming thermomechanical motion
  with microwave radiation pressure shot noise},\ }\href
  {https://doi.org/10.1103/PhysRevLett.116.013602} {\bibfield  {journal}
  {\bibinfo  {journal} {Phys. Rev. Lett.}\ }\textbf {\bibinfo {volume} {116}},\
  \bibinfo {pages} {013602} (\bibinfo {year} {2016})}\BibitemShut {NoStop}%
\bibitem [{\citenamefont {Genes}\ \emph {et~al.}(2008)\citenamefont {Genes},
  \citenamefont {Vitali}, \citenamefont {Tombesi}, \citenamefont {Gigan},\ and\
  \citenamefont
  {Aspelmeyer}}]{Genes2008-comparing-cold-damping-cavity-assisted}%
  \BibitemOpen
  \bibfield  {author} {\bibinfo {author} {\bibfnamefont {C.}~\bibnamefont
  {Genes}}, \bibinfo {author} {\bibfnamefont {D.}~\bibnamefont {Vitali}},
  \bibinfo {author} {\bibfnamefont {P.}~\bibnamefont {Tombesi}}, \bibinfo
  {author} {\bibfnamefont {S.}~\bibnamefont {Gigan}},\ and\ \bibinfo {author}
  {\bibfnamefont {M.}~\bibnamefont {Aspelmeyer}},\ }\bibfield  {title}
  {\bibinfo {title} {Ground-state cooling of a micromechanical oscillator:
  Comparing cold damping and cavity-assisted cooling schemes},\ }\href
  {https://doi.org/10.1103/PhysRevA.77.033804} {\bibfield  {journal} {\bibinfo
  {journal} {Phys. Rev. A}\ }\textbf {\bibinfo {volume} {77}},\ \bibinfo
  {pages} {033804} (\bibinfo {year} {2008})}\BibitemShut {NoStop}%
\bibitem [{\citenamefont {Lloyd}(2000)}]{Lloyd2000-coherent-quantum-feedback}%
  \BibitemOpen
  \bibfield  {author} {\bibinfo {author} {\bibfnamefont {S.}~\bibnamefont
  {Lloyd}},\ }\bibfield  {title} {\bibinfo {title} {Coherent quantum
  feedback},\ }\href {https://doi.org/10.1103/PhysRevA.62.022108} {\bibfield
  {journal} {\bibinfo  {journal} {Phys. Rev. A}\ }\textbf {\bibinfo {volume}
  {62}},\ \bibinfo {pages} {022108} (\bibinfo {year} {2000})}\BibitemShut
  {NoStop}%
\bibitem [{\citenamefont {Iida}\ \emph {et~al.}(2012)\citenamefont {Iida},
  \citenamefont {Yukawa}, \citenamefont {Yonezawa}, \citenamefont {Yamamoto},\
  and\ \citenamefont {Furusawa}}]{iida2012experimental}%
  \BibitemOpen
  \bibfield  {author} {\bibinfo {author} {\bibfnamefont {S.}~\bibnamefont
  {Iida}}, \bibinfo {author} {\bibfnamefont {M.}~\bibnamefont {Yukawa}},
  \bibinfo {author} {\bibfnamefont {H.}~\bibnamefont {Yonezawa}}, \bibinfo
  {author} {\bibfnamefont {N.}~\bibnamefont {Yamamoto}},\ and\ \bibinfo
  {author} {\bibfnamefont {A.}~\bibnamefont {Furusawa}},\ }\bibfield  {title}
  {\bibinfo {title} {Experimental demonstration of coherent feedback control on
  optical field squeezing},\ }\href
  {https://ieeexplore.ieee.org/stamp/stamp.jsp?arnumber=6189053} {\bibfield
  {journal} {\bibinfo  {journal} {IEEE Trans. Autom. Control}\ }\textbf
  {\bibinfo {volume} {57}},\ \bibinfo {pages} {2045} (\bibinfo {year}
  {2012})}\BibitemShut {NoStop}%
\bibitem [{\citenamefont {Nelson}\ \emph {et~al.}(2000)\citenamefont {Nelson},
  \citenamefont {Weinstein}, \citenamefont {Cory},\ and\ \citenamefont
  {Lloyd}}]{nelson2000experimental}%
  \BibitemOpen
  \bibfield  {author} {\bibinfo {author} {\bibfnamefont {R.~J.}\ \bibnamefont
  {Nelson}}, \bibinfo {author} {\bibfnamefont {Y.}~\bibnamefont {Weinstein}},
  \bibinfo {author} {\bibfnamefont {D.}~\bibnamefont {Cory}},\ and\ \bibinfo
  {author} {\bibfnamefont {S.}~\bibnamefont {Lloyd}},\ }\bibfield  {title}
  {\bibinfo {title} {Experimental demonstration of fully coherent quantum
  feedback},\ }\href
  {https://journals.aps.org/prl/abstract/10.1103/PhysRevLett.85.3045}
  {\bibfield  {journal} {\bibinfo  {journal} {Phys. Rev. Lett.}\ }\textbf
  {\bibinfo {volume} {85}},\ \bibinfo {pages} {3045} (\bibinfo {year}
  {2000})}\BibitemShut {NoStop}%
\bibitem [{\citenamefont {Karg}\ \emph {et~al.}(2020)\citenamefont {Karg},
  \citenamefont {Gouraud}, \citenamefont {Ngai}, \citenamefont {Schmid},
  \citenamefont {Hammerer},\ and\ \citenamefont
  {Treutlein}}]{Karg2020-Stron-coupling-membrane-atomic-spin}%
  \BibitemOpen
  \bibfield  {author} {\bibinfo {author} {\bibfnamefont {T.~M.}\ \bibnamefont
  {Karg}}, \bibinfo {author} {\bibfnamefont {B.}~\bibnamefont {Gouraud}},
  \bibinfo {author} {\bibfnamefont {C.~T.}\ \bibnamefont {Ngai}}, \bibinfo
  {author} {\bibfnamefont {G.-L.}\ \bibnamefont {Schmid}}, \bibinfo {author}
  {\bibfnamefont {K.}~\bibnamefont {Hammerer}},\ and\ \bibinfo {author}
  {\bibfnamefont {P.}~\bibnamefont {Treutlein}},\ }\bibfield  {title} {\bibinfo
  {title} {Light-mediated strong coupling between a mechanical oscillator and
  atomic spins 1 meter apart},\ }\href
  {https://doi.org/10.1126/science.abb0328} {\bibfield  {journal} {\bibinfo
  {journal} {Science}\ }\textbf {\bibinfo {volume} {369}},\ \bibinfo {pages}
  {174} (\bibinfo {year} {2020})}\BibitemShut {NoStop}%
\bibitem [{\citenamefont {Schmid}\ \emph {et~al.}(2022)\citenamefont {Schmid},
  \citenamefont {Ngai}, \citenamefont {Ernzer}, \citenamefont {Aguilera},
  \citenamefont {Karg},\ and\ \citenamefont
  {Treutlein}}]{Schmid2022-coherent-cooling-membrane-atomic-spin}%
  \BibitemOpen
  \bibfield  {author} {\bibinfo {author} {\bibfnamefont {G.-L.}\ \bibnamefont
  {Schmid}}, \bibinfo {author} {\bibfnamefont {C.~T.}\ \bibnamefont {Ngai}},
  \bibinfo {author} {\bibfnamefont {M.}~\bibnamefont {Ernzer}}, \bibinfo
  {author} {\bibfnamefont {M.~B.}\ \bibnamefont {Aguilera}}, \bibinfo {author}
  {\bibfnamefont {T.~M.}\ \bibnamefont {Karg}},\ and\ \bibinfo {author}
  {\bibfnamefont {P.}~\bibnamefont {Treutlein}},\ }\bibfield  {title} {\bibinfo
  {title} {Coherent feedback cooling of a nanomechanical membrane with atomic
  spins},\ }\href {https://doi.org/10.1103/PhysRevX.12.011020} {\bibfield
  {journal} {\bibinfo  {journal} {Phys. Rev. X}\ }\textbf {\bibinfo {volume}
  {12}},\ \bibinfo {pages} {011020} (\bibinfo {year} {2022})}\BibitemShut
  {NoStop}%
\bibitem [{\citenamefont {Hirose}\ and\ \citenamefont
  {Cappellaro}(2016)}]{Hirose2016-coherent-feedback-qubit-diamond}%
  \BibitemOpen
  \bibfield  {author} {\bibinfo {author} {\bibfnamefont {M.}~\bibnamefont
  {Hirose}}\ and\ \bibinfo {author} {\bibfnamefont {P.}~\bibnamefont
  {Cappellaro}},\ }\bibfield  {title} {\bibinfo {title} {Coherent feedback
  control of a single qubit in diamond},\ }\href
  {https://www.nature.com/articles/nature17404} {\bibfield  {journal} {\bibinfo
   {journal} {Nature}\ }\textbf {\bibinfo {volume} {532}},\ \bibinfo {pages}
  {77} (\bibinfo {year} {2016})}\BibitemShut {NoStop}%
\bibitem [{\citenamefont {Ernzer}\ \emph {et~al.}(2023)\citenamefont {Ernzer},
  \citenamefont {Bosch~Aguilera}, \citenamefont {Brunelli}, \citenamefont
  {Schmid}, \citenamefont {Karg}, \citenamefont {Bruder}, \citenamefont
  {Potts},\ and\ \citenamefont
  {Treutlein}}]{Ernzer2023-coherent-cooling-membrane}%
  \BibitemOpen
  \bibfield  {author} {\bibinfo {author} {\bibfnamefont {M.}~\bibnamefont
  {Ernzer}}, \bibinfo {author} {\bibfnamefont {M.}~\bibnamefont
  {Bosch~Aguilera}}, \bibinfo {author} {\bibfnamefont {M.}~\bibnamefont
  {Brunelli}}, \bibinfo {author} {\bibfnamefont {G.-L.}\ \bibnamefont
  {Schmid}}, \bibinfo {author} {\bibfnamefont {T.~M.}\ \bibnamefont {Karg}},
  \bibinfo {author} {\bibfnamefont {C.}~\bibnamefont {Bruder}}, \bibinfo
  {author} {\bibfnamefont {P.~P.}\ \bibnamefont {Potts}},\ and\ \bibinfo
  {author} {\bibfnamefont {P.}~\bibnamefont {Treutlein}},\ }\bibfield  {title}
  {\bibinfo {title} {Optical coherent feedback control of a mechanical
  oscillator},\ }\href {https://doi.org/10.1103/PhysRevX.13.021023} {\bibfield
  {journal} {\bibinfo  {journal} {Phys. Rev. X}\ }\textbf {\bibinfo {volume}
  {13}},\ \bibinfo {pages} {021023} (\bibinfo {year} {2023})}\BibitemShut
  {NoStop}%
\bibitem [{\citenamefont {Dania}\ \emph
  {et~al.}(2024{\natexlab{a}})\citenamefont {Dania}, \citenamefont {Bykov},
  \citenamefont {Goschin}, \citenamefont {Teller}, \citenamefont {Kassid},\
  and\ \citenamefont {Northup}}]{Dania2024-ultrahigh-quality-levitated}%
  \BibitemOpen
  \bibfield  {author} {\bibinfo {author} {\bibfnamefont {L.}~\bibnamefont
  {Dania}}, \bibinfo {author} {\bibfnamefont {D.~S.}\ \bibnamefont {Bykov}},
  \bibinfo {author} {\bibfnamefont {F.}~\bibnamefont {Goschin}}, \bibinfo
  {author} {\bibfnamefont {M.}~\bibnamefont {Teller}}, \bibinfo {author}
  {\bibfnamefont {A.}~\bibnamefont {Kassid}},\ and\ \bibinfo {author}
  {\bibfnamefont {T.~E.}\ \bibnamefont {Northup}},\ }\bibfield  {title}
  {\bibinfo {title} {Ultrahigh quality factor of a levitated nanomechanical
  oscillator},\ }\href {https://doi.org/10.1103/PhysRevLett.132.133602}
  {\bibfield  {journal} {\bibinfo  {journal} {Phys. Rev. Lett.}\ }\textbf
  {\bibinfo {volume} {132}},\ \bibinfo {pages} {133602} (\bibinfo {year}
  {2024}{\natexlab{a}})}\BibitemShut {NoStop}%
\bibitem [{\citenamefont {Rondin}\ \emph {et~al.}(2017)\citenamefont {Rondin},
  \citenamefont {Gieseler}, \citenamefont {Ricci}, \citenamefont {Quidant},
  \citenamefont {Dellago},\ and\ \citenamefont {Novotny}}]{rondin2017direct}%
  \BibitemOpen
  \bibfield  {author} {\bibinfo {author} {\bibfnamefont {L.}~\bibnamefont
  {Rondin}}, \bibinfo {author} {\bibfnamefont {J.}~\bibnamefont {Gieseler}},
  \bibinfo {author} {\bibfnamefont {F.}~\bibnamefont {Ricci}}, \bibinfo
  {author} {\bibfnamefont {R.}~\bibnamefont {Quidant}}, \bibinfo {author}
  {\bibfnamefont {C.}~\bibnamefont {Dellago}},\ and\ \bibinfo {author}
  {\bibfnamefont {L.}~\bibnamefont {Novotny}},\ }\bibfield  {title} {\bibinfo
  {title} {Direct measurement of kramers turnover with a levitated
  nanoparticle},\ }\href {https://www.nature.com/articles/nnano.2017.198}
  {\bibfield  {journal} {\bibinfo  {journal} {Nat. Nanotechnol.}\ }\textbf
  {\bibinfo {volume} {12}},\ \bibinfo {pages} {1130} (\bibinfo {year}
  {2017})}\BibitemShut {NoStop}%
\bibitem [{\citenamefont {Hebestreit}\ \emph {et~al.}(2018)\citenamefont
  {Hebestreit}, \citenamefont {Frimmer}, \citenamefont {Reimann},\ and\
  \citenamefont {Novotny}}]{hebestreit2018sensing}%
  \BibitemOpen
  \bibfield  {author} {\bibinfo {author} {\bibfnamefont {E.}~\bibnamefont
  {Hebestreit}}, \bibinfo {author} {\bibfnamefont {M.}~\bibnamefont {Frimmer}},
  \bibinfo {author} {\bibfnamefont {R.}~\bibnamefont {Reimann}},\ and\ \bibinfo
  {author} {\bibfnamefont {L.}~\bibnamefont {Novotny}},\ }\bibfield  {title}
  {\bibinfo {title} {Sensing static forces with free-falling nanoparticles},\
  }\href {https://doi.org/10.1103/PhysRevLett.121.063602} {\bibfield  {journal}
  {\bibinfo  {journal} {Phys. Rev. Lett.}\ }\textbf {\bibinfo {volume} {121}},\
  \bibinfo {pages} {063602} (\bibinfo {year} {2018})}\BibitemShut {NoStop}%
\bibitem [{\citenamefont {Conangla}\ \emph {et~al.}(2020)\citenamefont
  {Conangla}, \citenamefont {Rica},\ and\ \citenamefont
  {Quidant}}]{conangla2020extending}%
  \BibitemOpen
  \bibfield  {author} {\bibinfo {author} {\bibfnamefont {G.~P.}\ \bibnamefont
  {Conangla}}, \bibinfo {author} {\bibfnamefont {R.~A.}\ \bibnamefont {Rica}},\
  and\ \bibinfo {author} {\bibfnamefont {R.}~\bibnamefont {Quidant}},\
  }\bibfield  {title} {\bibinfo {title} {Extending vacuum trapping to absorbing
  objects with hybrid paul-optical traps},\ }\href
  {https://doi.org/https://doi.org/10.1021/acs.nanolett.0c02025} {\bibfield
  {journal} {\bibinfo  {journal} {Nano Lett.}\ }\textbf {\bibinfo {volume}
  {20}},\ \bibinfo {pages} {6018} (\bibinfo {year} {2020})}\BibitemShut
  {NoStop}%
\bibitem [{\citenamefont {Sun}\ \emph {et~al.}(2024)\citenamefont {Sun},
  \citenamefont {Pi}, \citenamefont {Kiang}, \citenamefont {Georgescu},
  \citenamefont {Ou}, \citenamefont {Ulbricht},\ and\ \citenamefont
  {Yan}}]{sun2024tunable}%
  \BibitemOpen
  \bibfield  {author} {\bibinfo {author} {\bibfnamefont {C.}~\bibnamefont
  {Sun}}, \bibinfo {author} {\bibfnamefont {H.}~\bibnamefont {Pi}}, \bibinfo
  {author} {\bibfnamefont {K.~S.}\ \bibnamefont {Kiang}}, \bibinfo {author}
  {\bibfnamefont {T.~S.}\ \bibnamefont {Georgescu}}, \bibinfo {author}
  {\bibfnamefont {J.-Y.}\ \bibnamefont {Ou}}, \bibinfo {author} {\bibfnamefont
  {H.}~\bibnamefont {Ulbricht}},\ and\ \bibinfo {author} {\bibfnamefont
  {J.}~\bibnamefont {Yan}},\ }\bibfield  {title} {\bibinfo {title} {Tunable
  on-chip optical traps for levitating particles based on single-layer
  metasurface},\ }\href
  {https://www.degruyterbrill.com/document/doi/10.1515/nanoph-2023-0873/html}
  {\bibfield  {journal} {\bibinfo  {journal} {Nanophotonics}\ }\textbf
  {\bibinfo {volume} {13}},\ \bibinfo {pages} {2791} (\bibinfo {year}
  {2024})}\BibitemShut {NoStop}%
\bibitem [{\citenamefont {Tomassi}\ \emph {et~al.}(2025)\citenamefont
  {Tomassi}, \citenamefont {Veldhuizen}, \citenamefont {Melo}, \citenamefont
  {Candoli}, \citenamefont {Riera-Campeny}, \citenamefont {Romero-Isart},
  \citenamefont {Meyer},\ and\ \citenamefont
  {Quidant}}]{Tomassi2025-inverted-potential}%
  \BibitemOpen
  \bibfield  {author} {\bibinfo {author} {\bibfnamefont {G.~F.~M.}\
  \bibnamefont {Tomassi}}, \bibinfo {author} {\bibfnamefont {D.}~\bibnamefont
  {Veldhuizen}}, \bibinfo {author} {\bibfnamefont {B.}~\bibnamefont {Melo}},
  \bibinfo {author} {\bibfnamefont {D.}~\bibnamefont {Candoli}}, \bibinfo
  {author} {\bibfnamefont {A.}~\bibnamefont {Riera-Campeny}}, \bibinfo {author}
  {\bibfnamefont {O.}~\bibnamefont {Romero-Isart}}, \bibinfo {author}
  {\bibfnamefont {N.}~\bibnamefont {Meyer}},\ and\ \bibinfo {author}
  {\bibfnamefont {R.}~\bibnamefont {Quidant}},\ }\bibfield  {title} {\bibinfo
  {title} {Accelerated state expansion of a nanoparticle in a dark inverted
  potential},\ }\href@noop {} {\  (\bibinfo {year} {2025})},\ \Eprint
  {https://arxiv.org/abs/2503.20707} {arXiv:2503.20707 [quant-ph]} \BibitemShut
  {NoStop}%
\bibitem [{\citenamefont {Magrini}\ \emph {et~al.}(2021)\citenamefont
  {Magrini}, \citenamefont {Rosenzweig}, \citenamefont {Bach}, \citenamefont
  {Deutschmann-Olek}, \citenamefont {Hofer}, \citenamefont {Hong},
  \citenamefont {Kiesel}, \citenamefont {Kugi},\ and\ \citenamefont
  {Aspelmeyer}}]{magrini2021real}%
  \BibitemOpen
  \bibfield  {author} {\bibinfo {author} {\bibfnamefont {L.}~\bibnamefont
  {Magrini}}, \bibinfo {author} {\bibfnamefont {P.}~\bibnamefont {Rosenzweig}},
  \bibinfo {author} {\bibfnamefont {C.}~\bibnamefont {Bach}}, \bibinfo {author}
  {\bibfnamefont {A.}~\bibnamefont {Deutschmann-Olek}}, \bibinfo {author}
  {\bibfnamefont {S.~G.}\ \bibnamefont {Hofer}}, \bibinfo {author}
  {\bibfnamefont {S.}~\bibnamefont {Hong}}, \bibinfo {author} {\bibfnamefont
  {N.}~\bibnamefont {Kiesel}}, \bibinfo {author} {\bibfnamefont
  {A.}~\bibnamefont {Kugi}},\ and\ \bibinfo {author} {\bibfnamefont
  {M.}~\bibnamefont {Aspelmeyer}},\ }\bibfield  {title} {\bibinfo {title}
  {Real-time optimal quantum control of mechanical motion at room
  temperature},\ }\href
  {https://doi.org/https://doi.org/10.1038/s41586-021-03602-3} {\bibfield
  {journal} {\bibinfo  {journal} {Nature}\ }\textbf {\bibinfo {volume} {595}},\
  \bibinfo {pages} {373} (\bibinfo {year} {2021})}\BibitemShut {NoStop}%
\bibitem [{\citenamefont {Tebbenjohanns}\ \emph {et~al.}(2021)\citenamefont
  {Tebbenjohanns}, \citenamefont {Mattana}, \citenamefont {Rossi},
  \citenamefont {Frimmer},\ and\ \citenamefont
  {Novotny}}]{tebbenjohanns2021quantum}%
  \BibitemOpen
  \bibfield  {author} {\bibinfo {author} {\bibfnamefont {F.}~\bibnamefont
  {Tebbenjohanns}}, \bibinfo {author} {\bibfnamefont {M.~L.}\ \bibnamefont
  {Mattana}}, \bibinfo {author} {\bibfnamefont {M.}~\bibnamefont {Rossi}},
  \bibinfo {author} {\bibfnamefont {M.}~\bibnamefont {Frimmer}},\ and\ \bibinfo
  {author} {\bibfnamefont {L.}~\bibnamefont {Novotny}},\ }\bibfield  {title}
  {\bibinfo {title} {Quantum control of a nanoparticle optically levitated in
  cryogenic free space},\ }\href
  {https://doi.org/https://doi.org/10.1038/s41586-021-03617-w} {\bibfield
  {journal} {\bibinfo  {journal} {Nature}\ }\textbf {\bibinfo {volume} {595}},\
  \bibinfo {pages} {378} (\bibinfo {year} {2021})}\BibitemShut {NoStop}%
\bibitem [{\citenamefont {Kamba}\ and\ \citenamefont
  {Aikawa}(2023)}]{Kamba2023-revealing-velocity-uncertainty}%
  \BibitemOpen
  \bibfield  {author} {\bibinfo {author} {\bibfnamefont {M.}~\bibnamefont
  {Kamba}}\ and\ \bibinfo {author} {\bibfnamefont {K.}~\bibnamefont {Aikawa}},\
  }\bibfield  {title} {\bibinfo {title} {Revealing the velocity uncertainties
  of a levitated particle in the quantum ground state},\ }\href
  {https://doi.org/10.1103/PhysRevLett.131.183602} {\bibfield  {journal}
  {\bibinfo  {journal} {Phys. Rev. Lett.}\ }\textbf {\bibinfo {volume} {131}},\
  \bibinfo {pages} {183602} (\bibinfo {year} {2023})}\BibitemShut {NoStop}%
\bibitem [{\citenamefont {Deli{\'c}}\ \emph {et~al.}(2020)\citenamefont
  {Deli{\'c}}, \citenamefont {Reisenbauer}, \citenamefont {Dare}, \citenamefont
  {Grass}, \citenamefont {Vuleti{\'c}}, \citenamefont {Kiesel},\ and\
  \citenamefont {Aspelmeyer}}]{delic2020cooling}%
  \BibitemOpen
  \bibfield  {author} {\bibinfo {author} {\bibfnamefont {U.}~\bibnamefont
  {Deli{\'c}}}, \bibinfo {author} {\bibfnamefont {M.}~\bibnamefont
  {Reisenbauer}}, \bibinfo {author} {\bibfnamefont {K.}~\bibnamefont {Dare}},
  \bibinfo {author} {\bibfnamefont {D.}~\bibnamefont {Grass}}, \bibinfo
  {author} {\bibfnamefont {V.}~\bibnamefont {Vuleti{\'c}}}, \bibinfo {author}
  {\bibfnamefont {N.}~\bibnamefont {Kiesel}},\ and\ \bibinfo {author}
  {\bibfnamefont {M.}~\bibnamefont {Aspelmeyer}},\ }\bibfield  {title}
  {\bibinfo {title} {Cooling of a levitated nanoparticle to the motional
  quantum ground state},\ }\href {https://doi.org/10.1126/science.aba3993}
  {\bibfield  {journal} {\bibinfo  {journal} {Science}\ }\textbf {\bibinfo
  {volume} {367}},\ \bibinfo {pages} {892} (\bibinfo {year}
  {2020})}\BibitemShut {NoStop}%
\bibitem [{\citenamefont {Piotrowski}\ \emph {et~al.}(2023)\citenamefont
  {Piotrowski}, \citenamefont {Windey}, \citenamefont {Vijayan}, \citenamefont
  {Gonzalez-Ballestero}, \citenamefont {de~los R{\'\i}os~Sommer}, \citenamefont
  {Meyer}, \citenamefont {Quidant}, \citenamefont {Romero-Isart}, \citenamefont
  {Reimann},\ and\ \citenamefont {Novotny}}]{piotrowski2022simultaneous}%
  \BibitemOpen
  \bibfield  {author} {\bibinfo {author} {\bibfnamefont {J.}~\bibnamefont
  {Piotrowski}}, \bibinfo {author} {\bibfnamefont {D.}~\bibnamefont {Windey}},
  \bibinfo {author} {\bibfnamefont {J.}~\bibnamefont {Vijayan}}, \bibinfo
  {author} {\bibfnamefont {C.}~\bibnamefont {Gonzalez-Ballestero}}, \bibinfo
  {author} {\bibfnamefont {A.}~\bibnamefont {de~los R{\'\i}os~Sommer}},
  \bibinfo {author} {\bibfnamefont {N.}~\bibnamefont {Meyer}}, \bibinfo
  {author} {\bibfnamefont {R.}~\bibnamefont {Quidant}}, \bibinfo {author}
  {\bibfnamefont {O.}~\bibnamefont {Romero-Isart}}, \bibinfo {author}
  {\bibfnamefont {R.}~\bibnamefont {Reimann}},\ and\ \bibinfo {author}
  {\bibfnamefont {L.}~\bibnamefont {Novotny}},\ }\bibfield  {title} {\bibinfo
  {title} {Simultaneous ground-state cooling of two mechanical modes of a
  levitated nanoparticle},\ }\href {https://doi.org/10.1038/s41567-023-01956-1}
  {\bibfield  {journal} {\bibinfo  {journal} {Nat. Phys.}\ } (\bibinfo {year}
  {2023})}\BibitemShut {NoStop}%
\bibitem [{\citenamefont {Ranfagni}\ \emph {et~al.}(2022)\citenamefont
  {Ranfagni}, \citenamefont {B\o{}rkje}, \citenamefont {Marino},\ and\
  \citenamefont {Marin}}]{Ranfagni2022-two-dimensional-quantum}%
  \BibitemOpen
  \bibfield  {author} {\bibinfo {author} {\bibfnamefont {A.}~\bibnamefont
  {Ranfagni}}, \bibinfo {author} {\bibfnamefont {K.}~\bibnamefont {B\o{}rkje}},
  \bibinfo {author} {\bibfnamefont {F.}~\bibnamefont {Marino}},\ and\ \bibinfo
  {author} {\bibfnamefont {F.}~\bibnamefont {Marin}},\ }\bibfield  {title}
  {\bibinfo {title} {Two-dimensional quantum motion of a levitated
  nanosphere},\ }\href {https://doi.org/10.1103/PhysRevResearch.4.033051}
  {\bibfield  {journal} {\bibinfo  {journal} {Phys. Rev. Res.}\ }\textbf
  {\bibinfo {volume} {4}},\ \bibinfo {pages} {033051} (\bibinfo {year}
  {2022})}\BibitemShut {NoStop}%
\bibitem [{\citenamefont {Dania}\ \emph
  {et~al.}(2024{\natexlab{b}})\citenamefont {Dania}, \citenamefont {Kremer},
  \citenamefont {Piotrowski}, \citenamefont {Candoli}, \citenamefont {Vijayan},
  \citenamefont {Romero-Isart}, \citenamefont {Gonzalez-Ballestero},
  \citenamefont {Novotny},\ and\ \citenamefont {Frimmer}}]{dania2024high}%
  \BibitemOpen
  \bibfield  {author} {\bibinfo {author} {\bibfnamefont {L.}~\bibnamefont
  {Dania}}, \bibinfo {author} {\bibfnamefont {O.~S.}\ \bibnamefont {Kremer}},
  \bibinfo {author} {\bibfnamefont {J.}~\bibnamefont {Piotrowski}}, \bibinfo
  {author} {\bibfnamefont {D.}~\bibnamefont {Candoli}}, \bibinfo {author}
  {\bibfnamefont {J.}~\bibnamefont {Vijayan}}, \bibinfo {author} {\bibfnamefont
  {O.}~\bibnamefont {Romero-Isart}}, \bibinfo {author} {\bibfnamefont
  {C.}~\bibnamefont {Gonzalez-Ballestero}}, \bibinfo {author} {\bibfnamefont
  {L.}~\bibnamefont {Novotny}},\ and\ \bibinfo {author} {\bibfnamefont
  {M.}~\bibnamefont {Frimmer}},\ }\bibfield  {title} {\bibinfo {title}
  {High-purity quantum optomechanics at room temperature},\ }\href@noop {} {\
  (\bibinfo {year} {2024}{\natexlab{b}})},\ \Eprint
  {https://arxiv.org/abs/2412.14117} {arXiv:2412.14117 [quant-ph]} \BibitemShut
  {NoStop}%
\bibitem [{\citenamefont {Conangla}\ \emph {et~al.}(2019)\citenamefont
  {Conangla}, \citenamefont {Ricci}, \citenamefont {Cuairan}, \citenamefont
  {Schell}, \citenamefont {Meyer},\ and\ \citenamefont
  {Quidant}}]{conangla2019optimal}%
  \BibitemOpen
  \bibfield  {author} {\bibinfo {author} {\bibfnamefont {G.~P.}\ \bibnamefont
  {Conangla}}, \bibinfo {author} {\bibfnamefont {F.}~\bibnamefont {Ricci}},
  \bibinfo {author} {\bibfnamefont {M.~T.}\ \bibnamefont {Cuairan}}, \bibinfo
  {author} {\bibfnamefont {A.~W.}\ \bibnamefont {Schell}}, \bibinfo {author}
  {\bibfnamefont {N.}~\bibnamefont {Meyer}},\ and\ \bibinfo {author}
  {\bibfnamefont {R.}~\bibnamefont {Quidant}},\ }\bibfield  {title} {\bibinfo
  {title} {Optimal feedback cooling of a charged levitated nanoparticle with
  adaptive control},\ }\href {https://doi.org/10.1103/PhysRevLett.122.223602}
  {\bibfield  {journal} {\bibinfo  {journal} {Phys. Rev. Lett.}\ }\textbf
  {\bibinfo {volume} {122}},\ \bibinfo {pages} {223602} (\bibinfo {year}
  {2019})}\BibitemShut {NoStop}%
\bibitem [{\citenamefont {Tebbenjohanns}\ \emph
  {et~al.}(2019{\natexlab{a}})\citenamefont {Tebbenjohanns}, \citenamefont
  {Frimmer}, \citenamefont {Militaru}, \citenamefont {Jain},\ and\
  \citenamefont {Novotny}}]{tebbenjohanns2019cold}%
  \BibitemOpen
  \bibfield  {author} {\bibinfo {author} {\bibfnamefont {F.}~\bibnamefont
  {Tebbenjohanns}}, \bibinfo {author} {\bibfnamefont {M.}~\bibnamefont
  {Frimmer}}, \bibinfo {author} {\bibfnamefont {A.}~\bibnamefont {Militaru}},
  \bibinfo {author} {\bibfnamefont {V.}~\bibnamefont {Jain}},\ and\ \bibinfo
  {author} {\bibfnamefont {L.}~\bibnamefont {Novotny}},\ }\bibfield  {title}
  {\bibinfo {title} {Cold damping of an optically levitated nanoparticle to
  microkelvin temperatures},\ }\href
  {https://doi.org/10.1103/PhysRevLett.122.223601} {\bibfield  {journal}
  {\bibinfo  {journal} {Phys. Rev. Lett.}\ }\textbf {\bibinfo {volume} {122}},\
  \bibinfo {pages} {223601} (\bibinfo {year} {2019}{\natexlab{a}})}\BibitemShut
  {NoStop}%
\bibitem [{\citenamefont {Vijayan}\ \emph {et~al.}(2023)\citenamefont
  {Vijayan}, \citenamefont {Zhang}, \citenamefont {Piotrowski}, \citenamefont
  {Windey}, \citenamefont {van~der Laan}, \citenamefont {Frimmer},\ and\
  \citenamefont {Novotny}}]{Vijayan2023-all-optical-scalable-CD}%
  \BibitemOpen
  \bibfield  {author} {\bibinfo {author} {\bibfnamefont {J.}~\bibnamefont
  {Vijayan}}, \bibinfo {author} {\bibfnamefont {Z.}~\bibnamefont {Zhang}},
  \bibinfo {author} {\bibfnamefont {J.}~\bibnamefont {Piotrowski}}, \bibinfo
  {author} {\bibfnamefont {D.}~\bibnamefont {Windey}}, \bibinfo {author}
  {\bibfnamefont {F.}~\bibnamefont {van~der Laan}}, \bibinfo {author}
  {\bibfnamefont {M.}~\bibnamefont {Frimmer}},\ and\ \bibinfo {author}
  {\bibfnamefont {L.}~\bibnamefont {Novotny}},\ }\bibfield  {title} {\bibinfo
  {title} {Scalable all-optical cold damping of levitated nanoparticles},\
  }\href {https://www.nature.com/articles/s41565-022-01254-6} {\bibfield
  {journal} {\bibinfo  {journal} {Nat. Nanotechnol.}\ }\textbf {\bibinfo
  {volume} {18}},\ \bibinfo {pages} {49} (\bibinfo {year} {2023})}\BibitemShut
  {NoStop}%
\bibitem [{SM()}]{SM}%
  \BibitemOpen
  \href@noop {} {\bibinfo {title} {See supplemental material for details on the
  theoretical description of the coherent feedback mechanism, the experimental
  setup and the data analysis.}}\BibitemShut {Stop}%
\bibitem [{\citenamefont {Gieseler}\ \emph {et~al.}(2012)\citenamefont
  {Gieseler}, \citenamefont {Deutsch}, \citenamefont {Quidant},\ and\
  \citenamefont {Novotny}}]{gieseler2012subkelvin}%
  \BibitemOpen
  \bibfield  {author} {\bibinfo {author} {\bibfnamefont {J.}~\bibnamefont
  {Gieseler}}, \bibinfo {author} {\bibfnamefont {B.}~\bibnamefont {Deutsch}},
  \bibinfo {author} {\bibfnamefont {R.}~\bibnamefont {Quidant}},\ and\ \bibinfo
  {author} {\bibfnamefont {L.}~\bibnamefont {Novotny}},\ }\bibfield  {title}
  {\bibinfo {title} {Subkelvin parametric feedback cooling of a laser-trapped
  nanoparticle},\ }\href
  {https://journals.aps.org/prl/abstract/10.1103/PhysRevLett.109.103603}
  {\bibfield  {journal} {\bibinfo  {journal} {Phys. Rev. Lett.}\ }\textbf
  {\bibinfo {volume} {109}},\ \bibinfo {pages} {103603} (\bibinfo {year}
  {2012})}\BibitemShut {NoStop}%
\bibitem [{\citenamefont {Jain}\ \emph {et~al.}(2016)\citenamefont {Jain},
  \citenamefont {Gieseler}, \citenamefont {Moritz}, \citenamefont {Dellago},
  \citenamefont {Quidant},\ and\ \citenamefont {Novotny}}]{jain2016direct}%
  \BibitemOpen
  \bibfield  {author} {\bibinfo {author} {\bibfnamefont {V.}~\bibnamefont
  {Jain}}, \bibinfo {author} {\bibfnamefont {J.}~\bibnamefont {Gieseler}},
  \bibinfo {author} {\bibfnamefont {C.}~\bibnamefont {Moritz}}, \bibinfo
  {author} {\bibfnamefont {C.}~\bibnamefont {Dellago}}, \bibinfo {author}
  {\bibfnamefont {R.}~\bibnamefont {Quidant}},\ and\ \bibinfo {author}
  {\bibfnamefont {L.}~\bibnamefont {Novotny}},\ }\bibfield  {title} {\bibinfo
  {title} {Direct measurement of photon recoil from a levitated nanoparticle},\
  }\href {https://doi.org/10.1103/PhysRevLett.116.243601} {\bibfield  {journal}
  {\bibinfo  {journal} {Phys. Rev. Lett.}\ }\textbf {\bibinfo {volume} {116}},\
  \bibinfo {pages} {243601} (\bibinfo {year} {2016})}\BibitemShut {NoStop}%
\bibitem [{\citenamefont {Tebbenjohanns}\ \emph
  {et~al.}(2019{\natexlab{b}})\citenamefont {Tebbenjohanns}, \citenamefont
  {Frimmer},\ and\ \citenamefont {Novotny}}]{tebbenjohanns2019optimal}%
  \BibitemOpen
  \bibfield  {author} {\bibinfo {author} {\bibfnamefont {F.}~\bibnamefont
  {Tebbenjohanns}}, \bibinfo {author} {\bibfnamefont {M.}~\bibnamefont
  {Frimmer}},\ and\ \bibinfo {author} {\bibfnamefont {L.}~\bibnamefont
  {Novotny}},\ }\bibfield  {title} {\bibinfo {title} {Optimal position
  detection of a dipolar scatterer in a focused field},\ }\href
  {https://doi.org/10.1103/PhysRevA.100.043821} {\bibfield  {journal} {\bibinfo
   {journal} {Phys. Rev. A}\ }\textbf {\bibinfo {volume} {100}},\ \bibinfo
  {pages} {043821} (\bibinfo {year} {2019}{\natexlab{b}})}\BibitemShut
  {NoStop}%
\bibitem [{\citenamefont {Kamba}\ \emph {et~al.}(2021)\citenamefont {Kamba},
  \citenamefont {Kiuchi}, \citenamefont {Yotsuya},\ and\ \citenamefont
  {Aikawa}}]{kamba2021recoil}%
  \BibitemOpen
  \bibfield  {author} {\bibinfo {author} {\bibfnamefont {M.}~\bibnamefont
  {Kamba}}, \bibinfo {author} {\bibfnamefont {H.}~\bibnamefont {Kiuchi}},
  \bibinfo {author} {\bibfnamefont {T.}~\bibnamefont {Yotsuya}},\ and\ \bibinfo
  {author} {\bibfnamefont {K.}~\bibnamefont {Aikawa}},\ }\bibfield  {title}
  {\bibinfo {title} {Recoil-limited feedback cooling of single nanoparticles
  near the ground state in an optical lattice},\ }\href
  {https://doi.org/10.1103/PhysRevA.103.L051701} {\bibfield  {journal}
  {\bibinfo  {journal} {Phys. Rev. A}\ }\textbf {\bibinfo {volume} {103}},\
  \bibinfo {pages} {L051701} (\bibinfo {year} {2021})}\BibitemShut {NoStop}%
\bibitem [{\citenamefont {Lindner}\ \emph {et~al.}(2024)\citenamefont
  {Lindner}, \citenamefont {Juschitz}, \citenamefont {Rieser}, \citenamefont
  {Fein}, \citenamefont {Debiossac}, \citenamefont {Ciampini}, \citenamefont
  {Aspelmeyer},\ and\ \citenamefont
  {Kiesel}}]{Lindner2024-hollow-core-ultra-high-vacuum}%
  \BibitemOpen
  \bibfield  {author} {\bibinfo {author} {\bibfnamefont {S.}~\bibnamefont
  {Lindner}}, \bibinfo {author} {\bibfnamefont {P.}~\bibnamefont {Juschitz}},
  \bibinfo {author} {\bibfnamefont {J.}~\bibnamefont {Rieser}}, \bibinfo
  {author} {\bibfnamefont {Y.~Y.}\ \bibnamefont {Fein}}, \bibinfo {author}
  {\bibfnamefont {M.}~\bibnamefont {Debiossac}}, \bibinfo {author}
  {\bibfnamefont {M.~A.}\ \bibnamefont {Ciampini}}, \bibinfo {author}
  {\bibfnamefont {M.}~\bibnamefont {Aspelmeyer}},\ and\ \bibinfo {author}
  {\bibfnamefont {N.}~\bibnamefont {Kiesel}},\ }\bibfield  {title} {\bibinfo
  {title} {Hollow-core fiber loading of nanoparticles into ultra-high vacuum},\
  }\href
  {https://pubs.aip.org/aip/apl/article/124/14/143501/3280323/Hollow-core-fiber-loading-of-nanoparticles-into}
  {\bibfield  {journal} {\bibinfo  {journal} {Appl. Phys. Lett.}\ }\textbf
  {\bibinfo {volume} {124}} (\bibinfo {year} {2024})}\BibitemShut {NoStop}%
\bibitem [{\citenamefont {Chao}\ \emph {et~al.}(2025)\citenamefont {Chao},
  \citenamefont {Hua}, \citenamefont {Liang}, \citenamefont {Yue},
  \citenamefont {Jia}, \citenamefont {You},\ and\ \citenamefont
  {Tey}}]{chao2025robust}%
  \BibitemOpen
  \bibfield  {author} {\bibinfo {author} {\bibfnamefont {Y.-X.}\ \bibnamefont
  {Chao}}, \bibinfo {author} {\bibfnamefont {Z.-X.}\ \bibnamefont {Hua}},
  \bibinfo {author} {\bibfnamefont {X.-H.}\ \bibnamefont {Liang}}, \bibinfo
  {author} {\bibfnamefont {Z.-P.}\ \bibnamefont {Yue}}, \bibinfo {author}
  {\bibfnamefont {C.}~\bibnamefont {Jia}}, \bibinfo {author} {\bibfnamefont
  {L.}~\bibnamefont {You}},\ and\ \bibinfo {author} {\bibfnamefont {M.~K.}\
  \bibnamefont {Tey}},\ }\bibfield  {title} {\bibinfo {title} {Robust
  suppression of high-frequency laser phase noise by adaptive pound-drever-hall
  feedforward},\ }\href
  {https://journals.aps.org/prapplied/abstract/10.1103/PhysRevApplied.23.L011005}
  {\bibfield  {journal} {\bibinfo  {journal} {Phys. Rev. Appl.}\ }\textbf
  {\bibinfo {volume} {23}},\ \bibinfo {pages} {L011005} (\bibinfo {year}
  {2025})}\BibitemShut {NoStop}%
\bibitem [{\citenamefont {Parniak}\ \emph {et~al.}(2021)\citenamefont
  {Parniak}, \citenamefont {Galinskiy}, \citenamefont {Zwettler},\ and\
  \citenamefont {Polzik}}]{parniak2021high}%
  \BibitemOpen
  \bibfield  {author} {\bibinfo {author} {\bibfnamefont {M.}~\bibnamefont
  {Parniak}}, \bibinfo {author} {\bibfnamefont {I.}~\bibnamefont {Galinskiy}},
  \bibinfo {author} {\bibfnamefont {T.}~\bibnamefont {Zwettler}},\ and\
  \bibinfo {author} {\bibfnamefont {E.~S.}\ \bibnamefont {Polzik}},\ }\bibfield
   {title} {\bibinfo {title} {High-frequency broadband laser phase noise
  cancellation using a delay line},\ }\href
  {https://opg.optica.org/oe/fulltext.cfm?uri=oe-29-5-6935&id=447872}
  {\bibfield  {journal} {\bibinfo  {journal} {Opt. Express}\ }\textbf {\bibinfo
  {volume} {29}},\ \bibinfo {pages} {6935} (\bibinfo {year}
  {2021})}\BibitemShut {NoStop}%
\bibitem [{\citenamefont {Pontin}\ \emph {et~al.}(2023)\citenamefont {Pontin},
  \citenamefont {Fu}, \citenamefont {Toro{\v{s}}}, \citenamefont {Monteiro},\
  and\ \citenamefont {Barker}}]{pontin2023simultaneous}%
  \BibitemOpen
  \bibfield  {author} {\bibinfo {author} {\bibfnamefont {A.}~\bibnamefont
  {Pontin}}, \bibinfo {author} {\bibfnamefont {H.}~\bibnamefont {Fu}}, \bibinfo
  {author} {\bibfnamefont {M.}~\bibnamefont {Toro{\v{s}}}}, \bibinfo {author}
  {\bibfnamefont {T.~S.}\ \bibnamefont {Monteiro}},\ and\ \bibinfo {author}
  {\bibfnamefont {P.~F.}\ \bibnamefont {Barker}},\ }\bibfield  {title}
  {\bibinfo {title} {Simultaneous cavity cooling of all six degrees of freedom
  of a levitated nanoparticle},\ }\href
  {https://www.nature.com/articles/s41567-023-02006-6} {\bibfield  {journal}
  {\bibinfo  {journal} {Nat. Phys.}\ }\textbf {\bibinfo {volume} {19}},\
  \bibinfo {pages} {1003} (\bibinfo {year} {2023})}\BibitemShut {NoStop}%
\bibitem [{\citenamefont {Gao}\ \emph {et~al.}(2024)\citenamefont {Gao},
  \citenamefont {van~der Laan}, \citenamefont {Zieli{\'n}ska}, \citenamefont
  {Militaru}, \citenamefont {Novotny},\ and\ \citenamefont
  {Frimmer}}]{gao2024feedback}%
  \BibitemOpen
  \bibfield  {author} {\bibinfo {author} {\bibfnamefont {J.}~\bibnamefont
  {Gao}}, \bibinfo {author} {\bibfnamefont {F.}~\bibnamefont {van~der Laan}},
  \bibinfo {author} {\bibfnamefont {J.~A.}\ \bibnamefont {Zieli{\'n}ska}},
  \bibinfo {author} {\bibfnamefont {A.}~\bibnamefont {Militaru}}, \bibinfo
  {author} {\bibfnamefont {L.}~\bibnamefont {Novotny}},\ and\ \bibinfo {author}
  {\bibfnamefont {M.}~\bibnamefont {Frimmer}},\ }\bibfield  {title} {\bibinfo
  {title} {Feedback cooling a levitated nanoparticle's libration to below 100
  phonons},\ }\href
  {https://journals.aps.org/prresearch/abstract/10.1103/PhysRevResearch.6.033009}
  {\bibfield  {journal} {\bibinfo  {journal} {Phys. Rev. Res.}\ }\textbf
  {\bibinfo {volume} {6}},\ \bibinfo {pages} {033009} (\bibinfo {year}
  {2024})}\BibitemShut {NoStop}%
\bibitem [{\citenamefont {Bang}\ \emph {et~al.}(2020)\citenamefont {Bang},
  \citenamefont {Seberson}, \citenamefont {Ju}, \citenamefont {Ahn},
  \citenamefont {Xu}, \citenamefont {Gao}, \citenamefont {Robicheaux},\ and\
  \citenamefont {Li}}]{bang2020five}%
  \BibitemOpen
  \bibfield  {author} {\bibinfo {author} {\bibfnamefont {J.}~\bibnamefont
  {Bang}}, \bibinfo {author} {\bibfnamefont {T.}~\bibnamefont {Seberson}},
  \bibinfo {author} {\bibfnamefont {P.}~\bibnamefont {Ju}}, \bibinfo {author}
  {\bibfnamefont {J.}~\bibnamefont {Ahn}}, \bibinfo {author} {\bibfnamefont
  {Z.}~\bibnamefont {Xu}}, \bibinfo {author} {\bibfnamefont {X.}~\bibnamefont
  {Gao}}, \bibinfo {author} {\bibfnamefont {F.}~\bibnamefont {Robicheaux}},\
  and\ \bibinfo {author} {\bibfnamefont {T.}~\bibnamefont {Li}},\ }\bibfield
  {title} {\bibinfo {title} {Five-dimensional cooling and nonlinear dynamics of
  an optically levitated nanodumbbell},\ }\href
  {https://doi.org/https://doi.org/10.1103/PhysRevResearch.2.043054} {\bibfield
   {journal} {\bibinfo  {journal} {Phys. Rev. Res.}\ }\textbf {\bibinfo
  {volume} {2}},\ \bibinfo {pages} {043054} (\bibinfo {year}
  {2020})}\BibitemShut {NoStop}%
\bibitem [{\citenamefont {Lepeshov}\ \emph {et~al.}(2023)\citenamefont
  {Lepeshov}, \citenamefont {Meyer}, \citenamefont {Maurer}, \citenamefont
  {Romero-Isart},\ and\ \citenamefont {Quidant}}]{lepeshov2023levitated}%
  \BibitemOpen
  \bibfield  {author} {\bibinfo {author} {\bibfnamefont {S.}~\bibnamefont
  {Lepeshov}}, \bibinfo {author} {\bibfnamefont {N.}~\bibnamefont {Meyer}},
  \bibinfo {author} {\bibfnamefont {P.}~\bibnamefont {Maurer}}, \bibinfo
  {author} {\bibfnamefont {O.}~\bibnamefont {Romero-Isart}},\ and\ \bibinfo
  {author} {\bibfnamefont {R.}~\bibnamefont {Quidant}},\ }\bibfield  {title}
  {\bibinfo {title} {Levitated optomechanics with meta-atoms},\ }\href
  {https://journals.aps.org/prl/abstract/10.1103/PhysRevLett.130.233601}
  {\bibfield  {journal} {\bibinfo  {journal} {Phys. Rev. Lett.}\ }\textbf
  {\bibinfo {volume} {130}},\ \bibinfo {pages} {233601} (\bibinfo {year}
  {2023})}\BibitemShut {NoStop}%
\bibitem [{\citenamefont {Afridi}\ \emph {et~al.}(2025)\citenamefont {Afridi},
  \citenamefont {Melo}, \citenamefont {Meyer},\ and\ \citenamefont
  {Quidant}}]{afridi2025controlling}%
  \BibitemOpen
  \bibfield  {author} {\bibinfo {author} {\bibfnamefont {A.}~\bibnamefont
  {Afridi}}, \bibinfo {author} {\bibfnamefont {B.}~\bibnamefont {Melo}},
  \bibinfo {author} {\bibfnamefont {N.}~\bibnamefont {Meyer}},\ and\ \bibinfo
  {author} {\bibfnamefont {R.}~\bibnamefont {Quidant}},\ }\bibfield  {title}
  {\bibinfo {title} {Controlling the sign of optical forces using metaoptics},\
  }\href@noop {} {\  (\bibinfo {year} {2025})},\ \Eprint
  {https://arxiv.org/abs/2504.18341} {arXiv:2504.18341 [physics]} \BibitemShut
  {NoStop}%
\bibitem [{\citenamefont {Lee}\ \emph {et~al.}(2025)\citenamefont {Lee},
  \citenamefont {Stickler}, \citenamefont {Pertsch},\ and\ \citenamefont
  {Hong}}]{lee2025inverse}%
  \BibitemOpen
  \bibfield  {author} {\bibinfo {author} {\bibfnamefont {M.}~\bibnamefont
  {Lee}}, \bibinfo {author} {\bibfnamefont {B.~A.}\ \bibnamefont {Stickler}},
  \bibinfo {author} {\bibfnamefont {T.}~\bibnamefont {Pertsch}},\ and\ \bibinfo
  {author} {\bibfnamefont {S.}~\bibnamefont {Hong}},\ }\bibfield  {title}
  {\bibinfo {title} {Inverse microparticle design for enhanced optical trapping
  and detection efficiency in all six degrees of freedom},\ }\href@noop {} {\
  (\bibinfo {year} {2025})},\ \Eprint {https://arxiv.org/abs/2506.01837}
  {arXiv:2506.01837 [physics]} \BibitemShut {NoStop}%
\bibitem [{\citenamefont {Dinter}\ \emph {et~al.}(2024)\citenamefont {Dinter},
  \citenamefont {Roberts}, \citenamefont {Volz}, \citenamefont {Schmidt},\ and\
  \citenamefont
  {Laplane}}]{Dinter2024-3d-displacement-sensing-mode-decomposition}%
  \BibitemOpen
  \bibfield  {author} {\bibinfo {author} {\bibfnamefont {T.}~\bibnamefont
  {Dinter}}, \bibinfo {author} {\bibfnamefont {R.}~\bibnamefont {Roberts}},
  \bibinfo {author} {\bibfnamefont {T.}~\bibnamefont {Volz}}, \bibinfo {author}
  {\bibfnamefont {M.~K.}\ \bibnamefont {Schmidt}},\ and\ \bibinfo {author}
  {\bibfnamefont {C.}~\bibnamefont {Laplane}},\ }\bibfield  {title} {\bibinfo
  {title} {Three-dimensional and selective displacement sensing of a levitated
  nanoparticle via spatial mode decomposition},\ }\href@noop {} {\  (\bibinfo
  {year} {2024})},\ \Eprint {https://arxiv.org/abs/2409.08827}
  {arXiv:2409.08827 [physics]} \BibitemShut {NoStop}%
\bibitem [{\citenamefont {Melo}\ \emph {et~al.}(2024)\citenamefont {Melo},
  \citenamefont {T~Cuairan}, \citenamefont {Tomassi}, \citenamefont {Meyer},\
  and\ \citenamefont {Quidant}}]{Melo2024-levitation-on-chip}%
  \BibitemOpen
  \bibfield  {author} {\bibinfo {author} {\bibfnamefont {B.}~\bibnamefont
  {Melo}}, \bibinfo {author} {\bibfnamefont {M.}~\bibnamefont {T~Cuairan}},
  \bibinfo {author} {\bibfnamefont {G.~F.~M.}\ \bibnamefont {Tomassi}},
  \bibinfo {author} {\bibfnamefont {N.}~\bibnamefont {Meyer}},\ and\ \bibinfo
  {author} {\bibfnamefont {R.}~\bibnamefont {Quidant}},\ }\bibfield  {title}
  {\bibinfo {title} {Vacuum levitation and motion control on chip},\ }\href
  {https://www.nature.com/articles/s41565-024-01677-3} {\bibfield  {journal}
  {\bibinfo  {journal} {Nat. Nanotechnol.}\ }\textbf {\bibinfo {volume} {19}},\
  \bibinfo {pages} {1270} (\bibinfo {year} {2024})}\BibitemShut {NoStop}%
\bibitem [{\citenamefont {Karg}\ \emph {et~al.}(2019)\citenamefont {Karg},
  \citenamefont {Gouraud}, \citenamefont {Treutlein},\ and\ \citenamefont
  {Hammerer}}]{Karg2019-remote-hamiltonian-mediated-light}%
  \BibitemOpen
  \bibfield  {author} {\bibinfo {author} {\bibfnamefont {T.~M.}\ \bibnamefont
  {Karg}}, \bibinfo {author} {\bibfnamefont {B.}~\bibnamefont {Gouraud}},
  \bibinfo {author} {\bibfnamefont {P.}~\bibnamefont {Treutlein}},\ and\
  \bibinfo {author} {\bibfnamefont {K.}~\bibnamefont {Hammerer}},\ }\bibfield
  {title} {\bibinfo {title} {Remote hamiltonian interactions mediated by
  light},\ }\href {https://doi.org/10.1103/PhysRevA.99.063829} {\bibfield
  {journal} {\bibinfo  {journal} {Phys. Rev. A}\ }\textbf {\bibinfo {volume}
  {99}},\ \bibinfo {pages} {063829} (\bibinfo {year} {2019})}\BibitemShut
  {NoStop}%
\bibitem [{\citenamefont {Metelmann}\ and\ \citenamefont
  {Clerk}(2015)}]{metelmann2015nonreciprocal}%
  \BibitemOpen
  \bibfield  {author} {\bibinfo {author} {\bibfnamefont {A.}~\bibnamefont
  {Metelmann}}\ and\ \bibinfo {author} {\bibfnamefont {A.~A.}\ \bibnamefont
  {Clerk}},\ }\bibfield  {title} {\bibinfo {title} {Nonreciprocal photon
  transmission and amplification via reservoir engineering},\ }\href
  {https://journals.aps.org/prx/abstract/10.1103/PhysRevX.5.021025} {\bibfield
  {journal} {\bibinfo  {journal} {Phys. Rev. X}\ }\textbf {\bibinfo {volume}
  {5}},\ \bibinfo {pages} {021025} (\bibinfo {year} {2015})}\BibitemShut
  {NoStop}%
\bibitem [{\citenamefont {Du}\ \emph {et~al.}(2025)\citenamefont {Du},
  \citenamefont {Monsel}, \citenamefont {Wieczorek},\ and\ \citenamefont
  {Splettstoesser}}]{du2025coherent}%
  \BibitemOpen
  \bibfield  {author} {\bibinfo {author} {\bibfnamefont {L.}~\bibnamefont
  {Du}}, \bibinfo {author} {\bibfnamefont {J.}~\bibnamefont {Monsel}}, \bibinfo
  {author} {\bibfnamefont {W.}~\bibnamefont {Wieczorek}},\ and\ \bibinfo
  {author} {\bibfnamefont {J.}~\bibnamefont {Splettstoesser}},\ }\bibfield
  {title} {\bibinfo {title} {Coherent feedback control for cavity
  optomechanical systems with a frequency-dependent mirror},\ }\href
  {https://journals.aps.org/pra/abstract/10.1103/PhysRevA.111.013506}
  {\bibfield  {journal} {\bibinfo  {journal} {Phys. Rev. A}\ }\textbf {\bibinfo
  {volume} {111}},\ \bibinfo {pages} {013506} (\bibinfo {year}
  {2025})}\BibitemShut {NoStop}%
\bibitem [{\citenamefont {Li}\ \emph {et~al.}(2017)\citenamefont {Li},
  \citenamefont {Li}, \citenamefont {Zippilli}, \citenamefont {Vitali},\ and\
  \citenamefont {Zhang}}]{li2017enhanced}%
  \BibitemOpen
  \bibfield  {author} {\bibinfo {author} {\bibfnamefont {J.}~\bibnamefont
  {Li}}, \bibinfo {author} {\bibfnamefont {G.}~\bibnamefont {Li}}, \bibinfo
  {author} {\bibfnamefont {S.}~\bibnamefont {Zippilli}}, \bibinfo {author}
  {\bibfnamefont {D.}~\bibnamefont {Vitali}},\ and\ \bibinfo {author}
  {\bibfnamefont {T.}~\bibnamefont {Zhang}},\ }\bibfield  {title} {\bibinfo
  {title} {Enhanced entanglement of two different mechanical resonators via
  coherent feedback},\ }\href
  {https://journals.aps.org/pra/abstract/10.1103/PhysRevA.95.043819} {\bibfield
   {journal} {\bibinfo  {journal} {Phys. Rev. A}\ }\textbf {\bibinfo {volume}
  {95}},\ \bibinfo {pages} {043819} (\bibinfo {year} {2017})}\BibitemShut
  {NoStop}%
\bibitem [{\citenamefont {Carlon~Zambon}\ \emph {et~al.}(2025)\citenamefont
  {Carlon~Zambon}, \citenamefont {Rossi}, \citenamefont {Frimmer},
  \citenamefont {Novotny}, \citenamefont {Gonzalez-Ballestero}, \citenamefont
  {Romero-Isart},\ and\ \citenamefont
  {Militaru}}]{Zambon2025-remote-entanglement-particles}%
  \BibitemOpen
  \bibfield  {author} {\bibinfo {author} {\bibfnamefont {N.}~\bibnamefont
  {Carlon~Zambon}}, \bibinfo {author} {\bibfnamefont {M.}~\bibnamefont
  {Rossi}}, \bibinfo {author} {\bibfnamefont {M.}~\bibnamefont {Frimmer}},
  \bibinfo {author} {\bibfnamefont {L.}~\bibnamefont {Novotny}}, \bibinfo
  {author} {\bibfnamefont {C.}~\bibnamefont {Gonzalez-Ballestero}}, \bibinfo
  {author} {\bibfnamefont {O.}~\bibnamefont {Romero-Isart}},\ and\ \bibinfo
  {author} {\bibfnamefont {A.}~\bibnamefont {Militaru}},\ }\bibfield  {title}
  {\bibinfo {title} {Motional entanglement of remote optically levitated
  nanoparticles},\ }\href {https://doi.org/10.1103/PhysRevA.111.013521}
  {\bibfield  {journal} {\bibinfo  {journal} {Phys. Rev. A}\ }\textbf {\bibinfo
  {volume} {111}},\ \bibinfo {pages} {013521} (\bibinfo {year}
  {2025})}\BibitemShut {NoStop}%
\end{thebibliography}%

\end{document}